\documentclass[10pt,fleqn]{article}

\usepackage{latexsym}
\usepackage{amsbsy}
\usepackage{amsfonts}
\usepackage{amssymb}
\usepackage{amscd}
\usepackage{t1enc}
\usepackage{epsfig}
\usepackage{color}
\usepackage{cite}

\usepackage{capt-of}
\usepackage{caption}

\def\appendix{\par
  \setcounter{section}{0}%
  \def\@chapapp{\appendixname}%
\def\thesection{Appendix \Alph{section}}}

\renewcommand{\appendixname}{Appendix}
\newtheorem{remark}{Remark}
\newtheorem{theorem}{Theorem}
\newtheorem{proposition}{Proposition}
\newtheorem{lemma}{Lemma}

\mathchardef\Gamma="0100
\mathchardef\Delta="0101
\mathchardef\Theta="0102
\mathchardef\Lambda="0103
\mathchardef\Xi="0104
\mathchardef\Pi="0105
\mathchardef\Sigma="0106
\mathchardef\Upsilon="0107
\mathchardef\Phi="0108
\mathchardef\Psi="0109
\mathchardef\Omega="010A

\newcommand{\sq}{\hbox{\rlap{$\sqcap$}$\sqcup$}}
\newcommand{\qed}{\ifmmode\sq\else{\unskip\nobreak\hfil
  \penalty50\hskip1em\null\nobreak\hfil\sq
  \parfillskip=0pt\finalhyphendemerits=0\endgraf}\fi{}}

\def\I{{\rm i}}

\def\D{{\rm d}}
\def\E{{\rm e}}

\def\vec{\boldsymbol}

\def\cal{\mathcal}

\def\rap{\!\!\!\! } 

\def\its{\it}

\def\primespe{\kern .08em '}
\def\basind#1{{\mbox{$_{#1}$}}} 

\input{diagxy}

\begin{document}

\title{Hyperbolic fractional-order Fourier transformations \\ in scalar theory of diffraction}

\date{}

\maketitle

\begin{center}

\vskip -1.2cm

{
\renewcommand{\thefootnote}{}
 {\bf   Pierre Pellat-Finet\footnote{\hskip -.53cm Laboratoire de Math\'ematiques de Bretagne Atlantique (LMBA) UMR CNRS 6205,

\noindent Universit\'e de Bretagne Sud, CS 60573, 56017 Vannes, France.

\noindent pierre.pellat-finet@univ-ubs.fr     
 }}
}
\setcounter{footnote}{0}

\medskip
    {\sl \small Universit\'e  Bretagne Sud,    UMR CNRS 6205, LMBA, F-56000 Vannes, France}

\smallskip 
\end{center}

\vskip .5cm 


\begin{center}
\begin{minipage}{12cm}

  \vskip -.2cm 
\hrulefill

\smallskip
{\small
  {\bf Abstract.} We define hyperbolic fractional-order Fourier transformations  by repla\-cing the  circular trigonometric functions   in the integral expressions of  conventional
  fractional-order Fourier transformations with hyperbolic trigonometric functions.  We establish the composition laws of  these hyperbolic transformations.  We then use hyperbolic fractional-order Fourier transforms  to mathematically represent Fresnel diffraction phenomena that cannot be described by conventional fractional Fourier transforms, due to  their  geometric configurations. Additionally, we apply appropriate compositions of  these transformations to coherent optical imaging.
  

\smallskip

\noindent {\bf Contents}

\smallskip
\ref{sect1}. Introduction: motivation\dotfill\pageref{sect1}

\ref{sect2}. Preliminary notions and results\dotfill\pageref{sect2}

\ref{sect3}. Hyperbolic fractional-order Fourier transformations\dotfill\pageref{sect3}

\hskip .4cm \ref{sect31}. Hyperbolic fractional-order Fourier transformations of the first kind\dotfill\pageref{sect31}

\hskip .4cm \ref{sect32}. Hyperbolic fractional-order Fourier transformations of the second  kind\dotfill\pageref{sect32}

\hskip .4cm\ref{sect33}.  Composition of two hyperbolic transformations of different kinds\dotfill\pageref{sect33}

\hskip .4cm\ref{sect34}. Algebra of hyperbolic fractional-order Fourier transformations\dotfill\pageref{sect34}

\ref{sect4}. Expressing diffraction by a hyperbolic fractional-order Fourier transformation\dotfill\pageref{sect4}

\hskip .4cm\ref{sect41}.  The case $J<-1$\dotfill\pageref{sect41}

\hskip .4cm\ref{sect42}.  The case $-1<J<0$\dotfill\pageref{sect42}

\ref{sect5}. Application to the refracting spherical cap\dotfill\pageref{sect5}

\hskip .4cm\ref{sect51}. The case $J<-1$ and $J'<-1$: product ${\cal H}_{\beta '}\circ{\cal H}_\beta$\dotfill\pageref{sect51}

\hskip .4cm\ref{sect52}. The case $-1<J<0$ and $-1<J'<0$: product ${\cal K}_{\beta '}\circ{\cal K}_\beta$\dotfill\pageref{sect52}

\ref{conc}. Conclusion\dotfill\pageref{conc}

\ref{appenA}. Proof of ${\cal H}_0={\cal I}$\dotfill\pageref{appenA}

References\dotfill\pageref{refe2}

\smallskip

}
\vskip -.1cm
\hrulefill
\end{minipage}
\end{center}

\section{Introduction: motivation}\label{sect1}

According to a scalar theory of diffraction, the optical-field transfer from a spherical emitter ${\cal A}$ (radius of curvature $R_A\ne 0$) to a receiver ${\cal B}$ (radius  of curvature $R_B\ne 0$) at a distance $D$ ($D\ne 0$) is expressed in the form \cite{PPF3,PPF1}
\begin{eqnarray}
  U_B(\vec r')\rap &=&\rap {\I\over \lambda D}\,\exp \left[-{\I\pi\over \lambda}\left({1\over R_B}+{1\over D}\right)r'^2\right] \nonumber \\
  & &\hskip 1cm \times \;\; \int_{{\mathbb R}^2}\exp \left[-{\I\pi\over \lambda}\left({1\over D}-{1\over R_A}\right)r^2\right]\,\exp\left({2\I\pi\over\lambda D} \vec r\vec\cdot\vec r'\right)\,U_A(\vec r)\,\D\vec r\,,\label{eq1}
  \end{eqnarray}
where $U_A$ denotes the field amplitude on ${\cal A}$, $U_B$ the field amplitude on ${\cal B}$, and $\lambda$ the radiation wavelength in the propagation medium (assumed to be homogeneous and isotropic). A constant phase factor $\exp (2\I\pi D/\lambda)$ has been omitted in Eq.\ (\ref{eq1}). Vectors $\vec r$ and $\vec r'$ are position vectors of generic points on ${\cal A}$ and ${\cal B}$ respectively (also called spatial variables): we have  $\vec r=(x,y)$, where $x$ and $y$ are orthogonal Cartesian coordinates (see Fig.\ \ref{fig1}).  We denote $r=\| \vec r\|=(x^2+y^2)^{1/2}$, and $\D\vec r=\D x\,\D y$.  The Euclidean scalar product  of $\vec r$ and $\vec r'$ is denoted $\vec r\vec\cdot\vec r'$. The norm $\|\vec r\|$ is physically homogeneous to a length (in SI, it is mesured in meters).  Finally, we point out that Eq.~(\ref{eq1}) is valid in the framework of the metaxial optics theory of G. Bonnet \cite{PPF1,GB1,GB2}, which constitutes a second-order approximation with respect to transverse dimensions of objects  and aperture angles (by comparison, paraxial or Gaussian optics constitutes a first-order approximation).

On the other hand, if $\alpha$ is a real number ($-\pi <0<\pi$), the two-dimensional fractional Fourier transform of order $\alpha$ of function $f$ is defined by (adapted from Namias \cite{Nam})
\begin{equation}
  {\cal F}_\alpha [f](\vec\rho\primespe)= {\I\,\E^{-\,\I\alpha}\over \sin\alpha}\exp(-\,\I\pi \rho\primespe^2\cot\alpha )
 \int_{{\mathbb R}^2}
  \exp(-\,\I\pi \rho^2\cot\alpha )\,\exp\left({2\I\pi \over \sin\alpha}\vec\rho \primespe\vec\cdot\vec\rho \right)\,f(\vec\rho)\,\D \vec\rho \,,\label{eq2}
\end{equation}
where $\vec \rho$ and $\vec \rho\primespe$ are two-dimensional vectors belonging to ${\mathbb R}^2$, with no physical dimensions.
In Equation ({\ref{eq2}), $f$ belongs to ${\cal S}({\mathbb R}^2)$, the vector space of rapidly decreasing functions. Fractional-order Fourier transformations can also be defined for complex orders \cite{Nam,Mcb}. They may be extended to ${\cal S}'({\mathbb R}^2)$, the space of tempered distributions: if $T$ is a tempered distribution, then ${\cal F}_\alpha [T]$ is defined for every $\varphi\in{\cal S}({\mathbb R}^2)$ by $\langle {\cal F}_\alpha[T],\varphi\rangle =\langle T,{\cal F}_\alpha[\varphi]\rangle$.

As well as the standard Fourier transformation does, fractional-order Fourier transformations give rise to an operational calculus, as explained by Namias \cite{Nam,Mcb}. 

The similarity between Eqs.\ (\ref{eq1}) and (\ref{eq2}) suggests that Eq.\ (\ref{eq1}) could be expressed through a fractional-order Fourier transformation, so that dealing with issues in diffraction might  benefit 
from the corresponding fractional operational calculus. This has been achieved and has  led to the development of ``fractional Fourier optics'' \cite{Oza,PPF3,PPF1}. A  method for doing so is as follows \cite{PPF4}.

We consider the diffraction-propagation phenomenon expressed by Eq.\ (\ref{eq1}) and introduce the parameter
\begin{equation}
  J={(R_A-D)(R_B+D)\over D(D-R_A+R_B)}\,.\label{eq3}\end{equation}
If $J>0$, we define $\alpha\in \,]-\pi, \pi [$ by
    \begin{equation}\cot^2\alpha =J\,,\hskip 1cm \alpha D>0\,,\hskip 1cm {DR_A\over R_A-D}\cot\alpha>0\,.\end{equation}
    For lightening the notation, we define
    \begin{equation}
      \varepsilon_A={D\over R_A-D}\cot\alpha\,,\hskip 1cm \varepsilon_B={D\over R_B+D}\cot\alpha\,,\end{equation}
    which are shown to be such that  $\varepsilon R_A>0$ and $\varepsilon_BR_B>0$.
    
    We then define reduced variables  $\vec \rho$ on ${\cal A}$ and $\vec \rho \primespe$ on ${\cal B}$
    by
    \begin{equation}
      \vec\rho ={\vec r\over\sqrt{\lambda\varepsilon_AR_A}}\,,\hskip 1cm  \vec\rho \primespe ={\vec r'\over\sqrt{\lambda\varepsilon_BR_B}}\,,\end{equation}
    and reduced field-amplitudes by
    \begin{equation}
      u_A(\vec \rho)=\sqrt{\lambda\varepsilon_AR_A}\,U_A\left(\sqrt{\lambda \varepsilon_AR_A}\,\vec \rho\right)\,,\hskip 1cm u_B(\vec \rho \primespe)=\sqrt{\lambda\varepsilon_BR_B}\,U_B\left(\sqrt{\lambda \varepsilon_BR_B}\,\vec\rho \primespe\right)\,,\end{equation}
    so that Eq.\ (\ref{eq1}) becomes \cite{PPF4}
    \begin{equation}
      u_B(\vec \rho\primespe)=\E^{\I\alpha}\,{\cal F}_\alpha [u_A](\vec \rho\primespe)\label{eq9n}\,.\end{equation}
 (Equation (\ref{eq9n}) remains valid if $\alpha =0$ and corresponds to  the limit of Eq.\ (\ref{eq1}) when $R_B=R_A$  and $D$ tends to $0$.)\goodbreak

    We conclude that, under the assumption $J>0$, the field transfer by diffraction from a spherical emitter to a spherical receiver can be expressed through a fractional-order Fourier transformation, the order of which depends on the geometrical configuration of the
    diffraction phenomenon (i.e. the distance from the  emitter to the receiver and their radii of curvature).

    The problem addressed in this article concerns the extension of the previous results to diffraction phenomena where $J<0$.
To solve this, we may first consider maintaining the equation $\cot^2\alpha =J$ by introducing complex $\alpha$,
 complex reduced variables and complex-order fractional Fourier transformations  \cite{PPF1,PPF10,PPF5}. We then change  complex variables to different real variables,  allowing us to introduce hyperbolic fractional Fourier transformations, as shown in a previous article \cite{PPF5}. In the present article, however, we propose an alternative solution, based on defining real orders $\beta$ such that $\coth^2\beta =-J$, or $\coth^2\beta =-1/J$, and directly introducing real reduced-variables and real-order hyperbolic fractional Fourier transformations, thus avoiding the use of complex orders and complex variables.

 \begin{remark} {\rm Since we will introduce hyperbolic transformations, we shall call ``circular'' the fractional-order Fourier transformations defined by Eq.\ (\ref{eq2}).}
 \end{remark}

 \begin{remark} {\rm A more appropriate definition would be to call the number $n=2\alpha/\pi$  the fractional order \cite{Nam}, so that the order of the standard Fourier transformation is equal to 1.  Fractional transformations with rational orders are then rational powers of the standard Fourier transformation.}
   \end{remark}

 \begin{remark} {\rm The standard Fourier transformation ${\cal F}$ is a special case of the circular fractional Fourier transformation: it corresponds to the order $\alpha =\pi /2$ (or $n=1$, according to the previous remark). Moreover, every circular transformation whose order is the product of $\pi /2$ and  a rational number corresponds to a rational power of the standard Fourier transformation. The name ``fractional Fourier transformation'' is thus justified---although ``fractional'' orders may be  the products of $\pi /2$ and real or complex numbers, not only rational ones. However, the standard Fourier transformation is not a special case of the hyperbolic fractional Fourier transformations, as we will define them (the hyperbolic transformation ${\cal K}_0$ is an exception, but fractional powers of ${\cal K}_0$ do not correspond to fractional orders). Calling these hyperbolic transformations ``fractional Fourier transformations'' may still be understandable, since they  correspond, in a sense, to circular fractional Fourier transformations with complex orders \cite{PPF5}. }
  \end{remark}

\section{Preliminary notions and results}\label{sect2}
    
\subsubsection*{\its Two-dimensional Fourier transformation} We define the two-dimensional Fourier transform of the rapidly decreasing function $f$ by
 \begin{equation}
   \widehat{f} (\vec\nu )=\int_{{\mathbb R}^2}\exp \bigl(2\I\pi \langle \vec\nu,\vec \rho\rangle\bigr)\,f(\vec \rho)\,\D \vec \rho\,,\label{eq9a}\end{equation}
 where $\vec \nu $ is the conjugate (or dual) variable of $\vec \rho$ and is an element of the  dual ${{\mathbb R}^2}^*$ of ${\mathbb R}^2$ ($\langle\vec \nu ,\vec\rho\rangle$ denotes the pairing of $\vec \nu$ and $\vec\rho$). We identify the dual ${{\mathbb R}^2}^*$ with ${\mathbb R}^2$  through the Euclidean scalar product: for every $\vec\nu \in {{\mathbb R}^2}^*$, there exists a unique $\vec\rho \primespe\in{\mathbb R}^2$ such that $\vec \rho \primespe\vec\cdot \vec \rho =\vec\nu (\vec\rho) = \langle \vec\nu ,\vec\rho\rangle$. Then we set $\vec\rho \primespe \equiv\vec\nu $, so that
 \begin{equation}
   \widehat f(\vec \rho \primespe)={\cal F}_{\pi /2}[f](\vec\rho \primespe)={\cal F}[f](\vec \rho\primespe)\,.\end{equation}

 \subsubsection*{\its A useful Fourier pair} If $\widehat f$ is the Fourier transform of $f$, we say that $f$ and $\widehat f$ form a Fourier pair, and
   we write $f\;\rightleftharpoons \;\widehat f$, or $f(\vec \rho)\;\rightleftharpoons \;\widehat f(\vec \nu )$. According to Eq.\ (\ref{eq9a}), if $A$ is a real number ($A\ne 0$), we have the Fourier  pair 
\begin{equation}
 \exp\left({\I\pi \rho^2\over A}\right)\;\rightleftharpoons \;\I A\exp (-\,\I\pi A\nu^2)\,,\label{eq11}
\end{equation}
where $\rho=\|\vec \rho\|$ and $\nu=\|\vec \nu \|$. Equation (\ref{eq11}) is valid for functions of a two-dimensional real variable, the Fourier transformation being defined by Eq.\ (\ref{eq9a}).

  \subsubsection*{\its Tempered distributions}
 The vector space of tempered distributions, denoted  ${\cal S}'({\mathbb R}^2)$, is the topological dual of ${\cal S}({\mathbb R}^2)$,   the vector space  of rapidly decreasing functions. If $T\in  {\cal S}'({\mathbb R}^2)$, its Fourier transform $\widehat{T}$ is defined by $ \langle \widehat T,\varphi\rangle = \langle T, \,\widehat{\!\varphi\!}\,\rangle,$ for every  $\varphi \in {\cal S}({\mathbb R}^2)$.

 The (two-dimensional) Dirac distribution, denoted $\delta$, is such that $\langle \delta ,\varphi\rangle =\varphi (\vec 0)$, for every $\varphi \in {\cal S}({\mathbb R}^2)$. Then
 \begin{equation}
   \langle \widehat \delta,\varphi\rangle = \langle \delta,\, \widehat {\!\varphi\!} \,\rangle
   =\widehat{\!\varphi\!}\,(\vec 0)=\int_{{\mathbb R}^2}\varphi (\vec \rho )\,\D\vec\rho =\langle 1,\varphi\rangle\,,\end{equation}
which gives $\widehat\delta =1$.

\subsubsection*{\its Parity operator and symmetrized function}

The symmetrized function $\widetilde f$ of a function $f$ is such that for every $\vec \rho$
\begin{equation}
  \widetilde f(\vec\rho )=f(-\vec\rho)\,.\end{equation}
The parity operator, denoted ${\cal P}$, is defined for every function $f$ by
${\cal P}[f](\vec \rho)=\widetilde f(\vec \rho)=f(-\vec\rho)$.

For every function $f$, we have  $\,\,\widehat{\!\!\widehat f}={\cal P}[f]=\widetilde f$,
that is,
$\,\,\widehat{\!\!\widehat f}(\vec\rho ) =\widetilde f(\vec\rho )=f(-\vec \rho)$.

\subsubsection*{\its Properties of  circular fractional-order Fourier transformations}
Circular fractional-order Fourier transformations are such that \cite{Nam,Mcb}:
\begin{itemize}
\item ${\cal F}_0={\cal I}$ (identity operator);
\item ${\cal F}_{\pi /2}={\cal F}$ (standard Fourier transformation);
\item ${\cal F}_{\alpha '}\circ{\cal F}_\alpha ={\cal F}_{\alpha '+\alpha}$ ;
  \item ${\cal F}_\alpha^{-1}={\cal F}_{-\alpha}$ ;
\item
  ${\cal F}_{\pi +\alpha}[f]={\cal F}_\alpha\bigl[\widetilde f\;\bigr]={\cal F}_\alpha\circ{\cal P}[f]={\cal P}\circ{\cal F}_\alpha[f]$ ;
\item ${\cal F}_{\pm\pi}={\cal P}$ (parity operator);
\item ${\cal F}_{\alpha +2n\pi}={\cal F}_{\alpha}$, $n\in {\mathbb Z}$ (extension of $\alpha\in [-\pi,\pi ]$  to $\alpha \in {\mathbb R}$).
\end{itemize}

\section{Hyperbolic fractional-order Fourier transformations}\label{sect3}

\subsection{Hyperbolic fractional-order Fourier transformations of the first kind}\label{sect31}

\subsubsection{Definition and properties}

  For $\beta \in{\mathbb R}$ ($\beta\ne 0$), we  denote ${\cal H}_\beta$  the two-dimensional hyperbolic fractional Fourier transformation of the first kind of order $\beta$, defined for $f$ in ${\cal S}({\mathbb R}^2)$ by
  \begin{equation}
  {\cal H}_{\beta} [f](\vec \rho \primespe)= {\I\,\E^{\,\beta}\over \sinh\beta}\exp(-\I\pi \rho\primespe^{\,2}\coth\beta )
 \!\int_{{\mathbb R}^2}\!\!\!
  \exp(-\I\pi \rho^2\coth\beta )\,\exp\!\left({2\I\pi \over \sinh\beta}\vec\rho\primespe\!\vec\cdot\vec\rho\!\right)f(\vec\rho)\,\D \vec\rho \,.\label{eq9}
  \end{equation}

  The integral in Eq.\ (\ref{eq9})
  may be seen as a Fourier transform. Indeed, if $f_\ddag$ denotes the function
  \begin{equation}
    f_\ddag : \vec\rho\;\longmapsto\;f_\ddag(\vec \rho )= \exp(-\I\pi \rho^2\coth\beta )\,f(\vec\rho)\,,\end{equation}
then
  \begin{equation}
    {\cal H}_{\beta} [f](\vec \rho \primespe)={\I\,\E^{\,\beta}\over \sinh\beta}\exp(-\I\pi \rho \primespe^{\, 2}\coth\beta )\;\widehat f_\ddag\left({\vec\rho \primespe\over\sinh\beta}\right)\,.\end{equation}\goodbreak
  We conclude that  hyperbolic fractional-order Fourier transformations inherit properties of the Fourier tranformation. For example,  ${\cal H}_\beta$ can be extended to tempered distributions, according to $\langle {\cal H}_\beta [T],\varphi \rangle =\langle T,{\cal H}_\beta [\varphi]\rangle$. They also inherit certains  properties of circular fractional-order Fourier transformations.

  We mention the following properties:
\begin{enumerate}
  \item[$\bullet$] When $\beta$ tends to 0,   ${\cal H}_\beta[f]\longrightarrow f$, for every function (or distribution) $f$, that is: ${\cal H}_0={\cal I}$ (identity operator). (A proof is given in \ref{appenA}.)
  \item[$\bullet$] ${\cal H}_{\beta '}[f]\longrightarrow {\cal H}_\beta [f]$ when $\beta '\longrightarrow\beta$.
    \item[$\bullet$] ${\cal H}_{\beta '}\circ {\cal H}_\beta={\cal H}_{\beta '+\beta}$.  (A proof is given in Section \ref{sect23}.)
    \item[$\bullet$]  ${{\cal H}_\beta}^{\!-1}={\cal H}_{-\beta}$. (A consequence of items above.)
      \item[$\bullet$] ${\cal H}_\beta\circ{\cal P}={\cal P}\circ{\cal H}_\beta$. (${\cal P}$ denotes the parity operator.)
  \end{enumerate}

\subsubsection{Some eigenfunctions}

\begin{proposition}\label{prop1}  The fonctions $g : \vec \rho \longmapsto \exp (-\,\I\pi\rho^2)$ and $h : \vec \rho\longmapsto \exp \I\pi \rho^2$ are eigenfunctions of ${\cal H}_\beta$, for every real number $\beta$, with respective eigenvalues $1$ and $\E^{2\beta}$, that is
    \begin{equation}
      {\cal H}_\beta[g]=g\,,\hskip .3cm \mbox{and}\;\;\;  {\cal H}_\beta[h]=\E^{2\beta}\,h\,.\end{equation}
\end{proposition}

 \noindent{\its Proof.}

 \medskip
 \noindent{\its i.} Since ${\cal H}_0= {\cal I}$, we have ${\cal H}_0[g]= g$ and  ${\cal H}_0[h]= h$, and Proposition 1 holds for $\beta =0$.

 \smallskip
 \noindent{\its ii.} For $\beta\ne 0$, we begin with $g$. The integral
\begin{eqnarray}
  I_1(\vec\rho\primespe)\rap &=&\rap  \int_{{\mathbb R}^2}\E^{-\,\I\pi \rho^2\coth\beta}
  \exp\left({2\I\pi \vec\rho \primespe\!\vec\cdot\vec\rho\over \sinh\beta}\right)
  \,\E^{-\,\I\pi \rho^2}\,\D \vec\rho\,,\nonumber \\
  &=& \rap \int_{{\mathbb R}^2}\E^{-\,\I\pi \rho^2(1+\coth\beta)}
  \exp\left({2\I\pi \vec\rho \primespe\!\vec\cdot\vec\rho \over \sinh\beta}\right)\,\D \vec \rho\,,
\end{eqnarray}
is a Fourier transform. 
We apply Eq.\ (\ref{eq11}) with $A=-1/(1+\coth\beta )=-\,\E^{-\beta}\,\sinh\beta\ne 0$, and we obtain
\begin{equation}
  \exp\left[-{\I\pi \rho^2(1+\coth\beta)}\right]\;\rightleftharpoons\, -\,\I\,\E^{-\beta}\,\sinh\beta \,\exp \bigl(\,\I\pi \nu^2\E^{-\beta}\sinh\beta \bigr).\label{eq12}
\end{equation}
The integral $I_1(\vec \rho \primespe)$ is the value  taken at $\vec \nu=\vec \rho \primespe/\sinh\beta$ by the function of the right side of Eq.\ (\ref{eq12}), namely
\begin{equation}
  I_1(\vec \rho\primespe)= -\,\I\,\E^{-\beta}\,\sinh\beta \,\exp \left(\I\pi \rho\primespe^{\, 2}\E^{-\beta}\over \sinh\beta\right)=
-\,\I\,\E^{-\beta}\,\sinh\beta \,\exp \bigl[\I\pi \rho\primespe^{\, 2}(\coth\beta -1)\bigr]\,.\label{eq13}
  \end{equation}
From Eqs. (\ref{eq9}) and (\ref{eq13}), for $g(\vec \rho)=\exp(- \,\I\pi \rho^2)$, we obtain
\begin{equation}
  {\cal H}_{\beta }[g](\vec \rho \primespe)=\exp (-\,\I\pi \rho \primespe^{\, 2})=g(\vec\rho \primespe)\,,\end{equation}
which means that the function $g$ is an eigenfunction of ${\cal H}_\beta$, with eigenvalue $1$.

\medskip
\noindent{\its iii.} For $\beta\ne 0$, and for $h(\vec \rho)=\exp \I\pi \rho^2$, we have the Fourier transform
\begin{eqnarray}
  I_2(\vec\rho\primespe)\rap &=&\rap  \int_{{\mathbb R}^2}\E^{-\,\I\pi \rho^2\coth\beta}
  \exp\left({2\I\pi \vec\rho\primespe\!\vec\cdot\vec\rho \over \sinh\beta}\right)
  \,\E^{\,\I\pi \rho^2}\,\D \vec\rho\,,\nonumber \\
  &=& \rap \int_{{\mathbb R}^2}\E^{\,\I\pi \rho^2(1-\coth\beta)}
  \exp\left({2\I\pi \vec\rho\primespe\!\vec\cdot\vec\rho\over \sinh\beta}\right)\,\D \rho\,.
\end{eqnarray}
 We use Eq.\ (\ref{eq11}) with
$A=1/(1-\coth\beta )=-\,\E^{\,\beta}\,\sinh\beta\ne 0$,
and obtain
\begin{equation}
  \exp\left[{\I\pi \rho^2(1-\coth\beta)}\right]\;\rightleftharpoons\, -\,\I\,\E^{\,\beta}\,\sinh\beta \,\exp (\I\pi \nu^2\E^{\,\beta}\sinh\beta)\,.\label{eq12b}
\end{equation}
The integral $I_2(\vec \rho \primespe)$ is the value  taken at $\vec \nu = \vec \rho \primespe/\sinh\beta$ by the function of the right side of Eq.\ (\ref{eq12b}), namely
\begin{equation}
  I_2(\vec \rho\primespe)= -\,\I\,\E^{\,\beta}\,\sinh\beta \,\exp \left(\I\pi \rho\primespe^{\, 2}\E^{\beta}\over \sinh\beta\right)=
-\I\,\E^{\beta}\,\sinh\beta \,\exp \bigl[\I\pi \rho\primespe^{\, 2}(\coth\beta +1)\bigr]\,.\label{eq13b}
  \end{equation}
From Eqs. (\ref{eq9}) and (\ref{eq13b}), for $h(\vec \rho)=\exp \I\pi \rho^2$, we obtain
\begin{equation}
  {\cal H}_{\beta }[h](\vec \rho \primespe)=\E^{2\beta}\exp (\I\pi \rho \primespe^{\, 2})=\E^{2\beta}\,h(\vec\rho \primespe)\,,\end{equation}
which means that the function $h$ is an eigenfunction of ${\cal H}_\beta$, with eigenvalue $\E^{2\beta}$.
\qed

\begin{remark} {\rm  Field transfers in  unstable optical resonators may be expressed by means of hyperbolic fractional Fourier transformations \cite{PPF5}. Eigenfunctions are then associated with the resonator propagating modes. For example, the function $h(\vec\rho )=\exp \I\pi \rho^2$ corresponds to a spherical wave, after the reduced variable $\vec \rho$ has been changed to the spatial variable $\vec r$ (see Sect.\ \ref{sect4} for the connection between spatial and reduced variables in diffraction theory).}
  \end{remark}

\subsubsection{Composition of two transformations of the first kind}\label{sect23}

\begin{proposition}\label{prop2}For every $\beta$ and $\beta '$ belonging to ${\mathbb R}$, we have ${\cal H}_{\beta '}\circ{\cal H}_{\beta}={\cal H}_{\beta '+\beta }$. The product is commutative.
\end{proposition}

\noindent{\its Proof.} Since ${\cal H}_0={\cal I}$, we have ${\cal H}_{0}\circ{\cal H}_\beta={\cal H}_{\beta }$ and ${\cal H}_{\beta '}\circ{\cal H}_0 ={\cal H}_{\beta '}$, and Proposition 2 holds for $\beta'\beta =0$.

If $\beta '+\beta =0$, since ${\cal H}_\beta^{-1}={\cal H}_{-\beta}={\cal H}_{\beta '}$, we obtain ${\cal H}_{\beta '}\circ{\cal H}_{\beta}={\cal H}_\beta^{-1}\circ{\cal H}_{\beta}={\cal I}={\cal H}_0={\cal H}_{\beta '+\beta }$ and Proposition 2 holds.

If $\beta '\beta\ne 0$ and $\beta '+\beta \ne 0$, the composition of the two transformations is given by
\begin{eqnarray}
{\cal H}_{\beta '}\circ{\cal H}_\beta [f](\vec\rho'')\rap &=&\rap {\cal H}_{\beta '}\bigl[{\cal H}_\beta [f]\bigr](\vec\rho '') \nonumber \\
\rap &=&\rap 
{\I\,\E^{\beta '}\over \sinh\beta '}\,\E^{-\I\pi \rho''^2\coth\beta '}\int_{{\mathbb R}^2}\E^{-\I\pi \rho\primespe^2\coth\beta '}\,\exp\left({2\I\pi \vec \rho\primespe\!\vec\cdot\vec\rho ''\over \sinh\beta '}\right)\,{\cal H}_\beta [f](\vec\rho ')\,\D \vec\rho '\nonumber \\
&=&\rap {-\E^{\beta +\beta '}\over \sinh\beta \sinh\beta '}\E^{-\I\pi \rho''^2\coth\beta '}\!\!\!
\int_{{\mathbb R}^2}\!\!\!\E^{-\I\pi \rho\primespe^2\coth\beta '}\exp\!\left({2\I\pi \vec\rho\primespe\!\vec\cdot \vec\rho ''\over \sinh\beta '}\right)\!\E^{-\I\pi \rho\primespe^2\coth\beta}\nonumber \\
& &\hskip .5cm \times \;\;\left\{\int_{{\mathbb R}^2}\!\E^{-\I\pi \rho^2\coth\beta}\,\exp\left({2\I\pi \vec\rho\primespe\!\vec\cdot\vec\rho \over \sinh\beta}\right)\,f(\vec\rho )\,\D \vec \rho\,\right\}\,\D \vec\rho\primespe\nonumber \\
\rap &=&\rap {-\E^{\beta +\beta '}\over \sinh\beta \sinh\beta '}\,\,\E^{-\I\pi \rho''^2\coth\beta '}\,\int_{{\mathbb R}^2}\!\!\E^{-\I\pi \rho^2\coth\beta }\,\left\{\int_{{\mathbb R}^2}\E^{-\I\pi \rho\primespe^2(\coth\beta +\coth\beta ')}\right.\nonumber \\
& &\hskip .5cm \times \;\left.\exp\!\left[2\I\pi \vec\rho \primespe\!\vec\cdot\left({\vec\rho''\over \sinh\beta '}+{\vec\rho\over \sinh\beta}\right)\right]\,\D \vec\rho \primespe\,\right\}\,f(\vec\rho)\,\D \vec\rho\,.
\end{eqnarray}

The integral between braces, denoted $I_3(\vec \rho'')$, is a Fourier transform. We use  Eq.\ (\ref{eq11}), once more, with
$A=-1/(\coth\beta +\coth\beta ')$, and obtain (for $\coth\beta +\coth\beta '\ne 0$, since $\beta '+\beta \ne 0$)
\begin{equation}
\E^{-\I\pi \rho\primespe^2(\coth\beta +\coth\beta ')} \,\rightleftharpoons\,
  {- \I\over \coth\beta +\coth\beta '} \,\exp\left({\I\pi \nu^2\over \coth\beta +\coth\beta '}\right)\,.\label{eq16}
\end{equation}

The integral $I_3(\vec \rho '')$ is the value of the function written on the right side of Eq.\ (\ref{eq16}) and taken at
$\vec \nu=(\vec \rho ''/ \sinh\beta ')+(\vec \rho /\sinh\beta)$, that is
\begin{eqnarray}
I_3(\vec \rho '')\rap&=&\rap\int_{{\mathbb R}^2}\!\E^{-\I\pi \rho\primespe^2(\coth\beta +\coth\beta ')}\,\exp \left[2\I\pi \vec\rho\primespe\!\vec\cdot\left({\vec \rho ''\over \sinh\beta '}+{\vec\rho \over \sinh\beta}\right)\right]\,\D \vec\rho\primespe  \\
&=&
{-\I\over \coth\beta +\coth\beta '}\exp\left({\I\pi \over \coth\beta +\coth\beta '}\,\left\|{\vec\rho ''\over \sinh\beta '}+{\vec\rho\over \sinh\beta}\right\|^2\right)\,.\nonumber 
\end{eqnarray}

We use
\begin{equation}
\sinh\beta\sinh\beta '(\coth\beta +\coth\beta ')=
\cosh\beta \sinh\beta '+\cosh\beta '\sinh\beta
=\sinh (\beta +\beta ')\,,\end{equation}
and we obtain
\begin{eqnarray}
{\cal H}_{\beta '}\circ{\cal H}_\beta [f](\vec\rho '')&=&
{\I\, \E^{(\beta +\beta ')}\over \sinh(\beta +\beta ')}
\, \E^{-\I\pi \rho''^2\coth\beta '} \,\int_{{\mathbb R}^2}\E^{-\I\pi \rho^2\coth\beta }    \label{eq19}
 \\
 & & \hskip 1cm \times \;\exp\left({\I\pi \over \coth\beta +\coth\beta '}\,\left\|{\vec\rho ''\over \sinh\beta '}+{\vec\rho\over \sinh\beta}\right\|^2\right)\,f(\vec\rho )\,\D \vec\rho\,.\nonumber 
\end{eqnarray}
Equation (\ref{eq19}) takes the form
\begin{equation}
{\cal H}_{\beta '}\circ{\cal H}_\beta [f](\vec\rho '')={\I\,\E^{\beta +\beta '}\over\sinh (\beta +\beta ')}
\,\E^{ \I\pi L\rho ''^2}\int_{{\mathbb R^2}}\E^{\I\pi M \rho^2} \E^{\I\pi N\vec\rho\vec\cdot\vec\rho''}f(\vec\rho)\,\D\vec\rho\,,
\end{equation}
where
\begin{equation}
N={2\over (\coth\beta +\coth\beta ')\sinh\beta\sinh\beta '} 
={2 \over \sinh (\beta +\beta ')}\,,
\end{equation}
\begin{eqnarray}
  M\rap&=&\rap -\coth\beta+{1\over (\coth\beta +\coth\beta ')\sinh^2\beta}\nonumber \\
  &=&\rap
  {-\cosh^2\beta -\cosh\beta\sinh\beta\coth\beta '+1\over (\coth\beta +\coth\beta ')\sinh^2\beta }\nonumber \\
  &=&\rap
  {-\sinh\beta -\cosh\beta\coth\beta '\over (\coth\beta +\coth\beta ')\sinh\beta }\nonumber \\
  &=& \rap
  {-1 -\coth\beta\coth\beta '\over \coth\beta +\coth\beta ' } \nonumber \\
  &=&\rap
  -\coth (\beta +\beta ')\,.
\end{eqnarray}
A similar derivation leads to $L=-\coth (\beta +\beta ')$.

Eventually, we obtain
\begin{eqnarray}
{\cal H}_{\beta '}\circ{\cal H}_\beta [f](\vec\rho'')\rap &=&\rap 
{\I\E^{\beta +\beta '}\over \sinh(\beta +\beta ')}
\, \E^{-\I\pi \rho''^2 \coth(\beta + \beta ')}\nonumber \\
& &\hskip 2cm \times \;\;\int_{{\mathbb R}^2}\E^{-\I\pi \rho^2\coth (\beta + \beta ' )}\exp\left({2\I\pi \vec\rho\vec\cdot\vec\rho ''\over \sinh(\beta +\beta ')}\right)\,f(\vec\rho)\,\D \vec\rho \nonumber \\
\rap &=&\rap {\cal H}_{\beta +\beta '}[f](\vec\rho'')\,.
\end{eqnarray}
The proof is complete.\qed

\subsection{Hyperbolic fractional-order Fourier transformations of the second  kind}\label{sect32}

\subsubsection{Definition}

  For $\beta \in{\mathbb R}$, we  denote ${\cal K}_\beta$  the two-dimensional hyperbolic fractional Fourier transformation of the second kind of order $\beta$, defined for every function $f$ in ${\cal S}({\mathbb R}^2)$ by
  \begin{equation}
  {\cal K}_{\beta} [f](\vec \rho \primespe)= {\I\,\E^{\beta}\over \cosh\beta}\exp(\I\pi \rho \primespe^{\, 2}\tanh\beta )
 \!\int_{{\mathbb R}^2}\!\!
  \exp(-\I\pi \rho^2\tanh\beta )\,\exp\left({2\I\pi \over \cosh\beta}\vec\rho \primespe\!\vec\cdot\vec\rho\right)f(\vec\rho)\,\D \vec\rho \,.\label{eq85}
  \end{equation}

  \goodbreak
Properties of $K_\beta$:
  \begin{enumerate}
  \item[$\bullet$] ${\cal K}_0[f]=\I\,\widehat f$ (i.e. ${\cal K}_0=\I\,{\cal F}$) for every function (or distribution) $f$; then ${\cal K}_0\ne {\cal I}$ (identity operator).
     \item[$\bullet$] ${{\cal K}_0}^2=-{\cal P}$ ; ${{\cal K}_0}^4={\cal I}$.
    \item[$\bullet$] ${\cal K}_\beta\circ{\cal P}={\cal P}\circ{\cal K}_\beta$.
  \end{enumerate}

  The composition of two hyperbolic fractional-order Fourier transformations is examined in Section \ref{sect223}.

\subsubsection{Some eigenfunctions}

\begin{lemma}\label{lem1} Let $g$ and $h$ be defined by $g(\vec \rho)=\exp(-\I\pi \rho^2)$ and $h(\vec \rho)=\exp\I\pi \rho^2$. Then, for every $\beta\in{\mathbb R}$, we have  $ {\cal K}_\beta[g]=h$, and ${\cal K}_\beta[h]=-\E^{2\beta}g$.
 \end{lemma}
  
  \noindent{\its Proof.}
  
  \noindent {\its i.}
We consider the integral
\begin{eqnarray}
  I_4(\vec\rho\primespe)\rap &=&\rap \int_{\mathbb R}\E^{-\I\pi \rho^2\tanh\beta}
  \exp\left({2\I\pi \vec\rho\primespe\:\vec\cdot\vec\rho \over \cosh\beta}\right)
  \,\E^{-\I\pi \rho^2}\,\D \vec\rho\,, \nonumber \\
  &=&\rap  \int_{\mathbb R}\E^{-\I\pi \rho^2(\tanh\beta +1)}
  \exp\left({2\I\pi  \vec\rho\primespe\!\vec\cdot\vec\rho \over \cosh\beta}\right)\,\D \vec\rho\,,
\end{eqnarray}
which is  a Fourier transform.
We apply Eq.\ (\ref{eq11}) with $A=-1/(\tanh\beta +1)=-\E^{-\beta}\cosh\beta\ne 0$, and obtain
\begin{equation}
  \exp\left[-{\I\pi \rho^2(\tanh\beta +1)}\right]\;\rightleftharpoons\, -\I\, \E^{-\beta}\cosh\beta\,\exp\left(\I\pi \nu^2\,\E^{-\beta}\cosh\beta\right).\label{eq87}
\end{equation}
The integral $I_4(\vec\rho\primespe)$ is the value of the function on the right side, taken at $\vec \nu=\vec \rho '/\cosh\beta$, that is
\begin{equation}
  I_4(\vec\rho \primespe)=-\I\,\E^{-\beta}\cosh\beta\,\exp\left({\I\pi \E^{-\beta}\rho\primespe^{\,2}\over\cosh\beta}\right)
  =-\I\,\E^{-\beta}\cosh\beta\,\exp\left[\I\pi\rho\primespe^{\, 2}(1-\tanh\beta)\right]\,.
\end{equation}
We conclude with
\begin{equation}
  {\cal K}_\beta[g](\vec\rho \primespe)={\I\,\E^{\, \beta}\over \cosh\beta}\exp(\I\pi \rho\primespe^{\, 2}\tanh\beta)\,I_4 (\vec\rho\primespe)=\exp(\I\pi\rho\primespe^{\, 2})=h(\vec\rho \primespe)\,.
\end{equation}

\medskip
\noindent {\its ii.} We consider
\begin{eqnarray}
  I_5(\vec\rho\primespe)\rap &=&\rap  \int_{\mathbb R}\E^{-\I\pi \rho^2\tanh\beta}
  \exp\left({2\I\pi \vec\rho\primespe\!\vec\cdot\vec\rho \over \cosh\beta}\right)
  \,\E^{\I\pi \rho^2}\,\D \vec\rho\,,\nonumber \\
  &=&\rap \int_{\mathbb R}\E^{-\I\pi \rho^2(\tanh\beta -1)}
  \exp\left({2\I\pi  \vec\rho\primespe\!\vec\cdot\vec\rho\over \cosh\beta}\right)\,\D \vec\rho\,.
\end{eqnarray}
We apply Eq.\ (\ref{eq11}) with $A=-1/(\tanh\beta -1)=\E^{\beta}\cosh\beta\ne 0$ and we obtain
\begin{equation}
  \exp\left[-{\I\pi \rho^2(\tanh\beta -1)}\right]\;\rightleftharpoons\, \I\, \E^{\beta}\cosh\beta\,\exp\left(-\I\pi \nu^2\,\E^{\beta}\cosh\beta\right)\,.
\end{equation}
Then
\begin{equation}
  I_5(\vec\rho \primespe)=\I\,\E^{\beta}\cosh\beta\,\exp\left(-{\I\pi \E^{\beta}\rho\primespe^{\,2}\over\cosh\beta}\right)
  =\I\,\E^{\beta}\cosh\beta\,\exp\left[-\I\pi\rho\primespe^{\, 2}(1+\tanh\beta)\right]\,.
\end{equation}
We conclude with
\begin{equation}
  {\cal K}_\beta[h](\vec\rho \primespe)={\I\,\E^{\, \beta}\over \cosh\beta}\exp(\I\pi \rho\primespe^{\, 2}\tanh\beta)\,I_5 (\vec\rho\primespe)=-\E^{2\beta}\exp(-\I\pi\rho\primespe^{\, 2})=-\E^{2\beta}g(\vec\rho \primespe)\,.
\end{equation}
The proof is complete.
\qed

\begin{proposition}\label{prop3} Let $g$ and $h$ be as in Lemma \ref{lem1}. For every $\beta\in{\mathbb R}$, the function $f_{_-}$ defined by
  \begin{eqnarray}
    f_{_-}(\vec \rho)\rap &=&\rap (1-\I)\E^{\beta /2}g(\vec \rho)+ (1+\I)\E^{-\beta /2}h(\vec\rho )\nonumber \\ \rap &=&\rap  (1-\I)\E^{\beta /2}\exp(-\I\pi\rho^2) + (1+\I)\E^{-\beta /2}\exp (\I\pi\rho^2)\,,
  \end{eqnarray}
  is an eigenfunction of ${\cal K}_\beta$ with eigenvalue $-\I\,\E^{\beta}$.
The function  $f_{_+}$ defined by
  \begin{eqnarray}
    f_{_+}(\vec \rho)\rap &=&\rap (1-\I)\E^{\beta /2}g(\vec \rho)- (1+\I)\E^{-\beta /2}h(\vec\rho )\nonumber \\ \rap &=&\rap  (1-\I)\E^{\beta /2}\exp(-\I\pi\rho^2) - (1+\I)\E^{-\beta /2}\exp (\I\pi\rho^2)\,,
  \end{eqnarray}
  is an eigenfunction with eigenvalue $\I\,\E^{\beta}$.
 \end{proposition}

\noindent{\its Proof.}
A simple checking would prove the proposition. Nevertheless, we indicate how to derive the result without guessing it a priori. 
We look for an eigenfunction of ${\cal K}_\beta$ as a linear combination of functions $g$ and $h$, that is $f=ag+bf$ (with $ab\ne 0$). The function $f$ is an eigenfunction of ${\cal K}_\beta$ if there is a number $\Lambda$ (not zero) such that
\begin{equation}
  \Lambda f={\cal K}_\beta[f]={\cal K}_\beta[ag+bh]\,.
\end{equation}
According to Lemma \ref{lem1}, we have
\begin{equation}
  {\cal K}_\beta[ag+bh]=a{\cal K}_\beta[g]+b{\cal K}_\beta[h]=ah-b\E^{2\beta}g\,.
\end{equation}
For $f$ to be an eigenfunction, we must have $\Lambda ag+\Lambda bh=\Lambda f= ah-b\,\E^{2\beta}g$, and since the functions $g$ and $h$ are linearly independent, we must have
\begin{equation}
  \Lambda a=-b\,\E^{2\beta}\,,\hskip .5cm \mbox{and}\;\;\;\Lambda b=a\,.\end{equation}
We deduce $\Lambda^2b=-b\,\E^{2\beta}$, so that $\Lambda =\pm\,\I \,\E^{\beta}$.
Since eigenfunctions are defined up to a multiplicative factor, we choose $a=1$, and from $\Lambda b=a=1$, we obtain
$b=\mp\,\I\,\E^{-\beta}$, that is, for $\Lambda =\I\E^{\beta}$
\begin{equation}
  f=g-\I\,\E^{-\beta} h\,,\end{equation}
and for  $\Lambda =-\I\,\E^{\beta}$
\begin{equation}
  f=g+\I\,\E^{-\beta} h\,.\end{equation}
More symmetric forms for $f$ would be
\begin{equation}
  f_{_+}=(1-\I)\E^{\beta /2}g-(1+\I)\E^{-\beta/2}h\,,\hskip .5cm \mbox{for}\;\;\; \Lambda =\I\,\E^{\beta} \,,\end{equation}
and
\begin{equation}
  f_{_-}=(1-\I)\E^{\beta /2}g+(1+\I)\E^{-\beta/2}h\,,\hskip .5cm \mbox{for}\;\;\; \Lambda =-\I\,\E^{\beta} \,.\end{equation}

We eventually check
\begin{eqnarray}
  {\cal K}_\beta[f_{_+}]\rap &=&\rap {\cal K}_\beta\left[(1-\I)\E^{\beta /2}g-(1+\I)\E^{-\beta/2} h\right]\nonumber\\
  \rap &=&\rap (1-\I)\E^{\beta /2}h+ (1+\I)\E^{3\beta /2}g \nonumber\\
    \rap &=&\rap \I\,\E^{\beta}\bigl[(1-\I)\E^{\beta /2}g-(1+\I)\E^{-\beta/2} h\bigr]\nonumber\\
    \rap &=&\rap \I\,\E^{\beta}f_{_+}\,.
\end{eqnarray}
and
\begin{eqnarray}
  {\cal K}_\beta[f_{_-}]\rap &=&\rap {\cal K}_\beta\left[(1-\I)\E^{\beta /2}g+(1+\I)\E^{-\beta/2} h\right]\nonumber\\
  \rap &=&\rap (1-\I)\E^{\beta /2}h- (1+\I)\E^{3\beta /2}g\nonumber\\
    \rap &=&\rap -\I\E^{\beta}\bigl[(1-\I)\E^{\beta /2}g+(1+\I)\E^{-\beta/2} h\bigr]\nonumber\\
    \rap &=&\rap -\I\,\E^{\beta}f_{_-}\,.
\end{eqnarray}
The  proof is complete. \qed

\subsubsection{Composition of two transformations of the second kind}\label{sect223}

\begin{proposition}\label{prop4}For every $\beta$ and $\beta '$ belonging to ${\mathbb R}$, we have ${\cal K}_{\beta '}\circ{\cal K}_{\beta}=-\E^{2\beta '}\,{\cal P}\circ {\cal H}_{\beta -\beta '}$, where ${\cal P}$ denotes the parity operator. The product of two hyperbolic fractional-order Fourier transformations of the second kind is not commutative.
\end{proposition}

\noindent{\its Proof.}

\noindent {\its i.}
The composition of the two transformations ${\cal K}_\beta$ and ${\cal K}_{\beta '}$ is given by
\begin{eqnarray}
{\cal K}_{\beta '}\circ{\cal K}_\beta [f](\vec\rho'')\rap &=&\rap {\cal K}_{\beta '}\bigl[{\cal K}_\beta [f]\bigr](\vec\rho '') \label{eq57}\\
\rap &=&\rap 
{\I\;\E^{\beta '}\over \cosh\beta '}\,\E^{\I\pi \rho''^2\tanh\beta '}\int_{{\mathbb R}^2}\E^{-\I\pi \rho\primespe^2\tanh\beta '}\,\exp\left({2\I\pi \vec \rho \primespe\!\vec\cdot\vec\rho ''\over \cosh\beta '}\right)\,{\cal K}_\beta [f](\vec\rho \primespe)\,\D \vec\rho \primespe\nonumber \\
&=&\rap {-\E^{\beta +\beta '}\over \cosh\beta \cosh\beta '}\,\E^{\I\pi \rho''^2\tanh\beta '}\!\!
\int_{{\mathbb R}^2}\!\!\E^{-\I\pi \rho\primespe^2\tanh\beta '}\exp\left({2\I\pi \vec\rho\primespe\!\vec\cdot \vec\rho ''\over \cosh\beta '}\right)\, \E^{\I\pi \rho'^2\tanh\beta}\nonumber \\
& &\hskip .5cm \times \;\;\left\{\int_{{\mathbb R}^2}\E^{-\I\pi \rho^2\tanh\beta}\,\exp\left({2\I\pi \vec\rho\primespe\!\vec\cdot\vec\rho \over \cosh\beta}\right)\,f(\vec\rho )\,\D \vec \rho\,\right\}\,\D \vec\rho \primespe\nonumber \\
\rap &=&\rap {-\E^{\beta +\beta '}\over \cosh\beta \cosh\beta '}\,\,\E^{\I\pi \rho''^2\tanh\beta '}\,\int_{{\mathbb R}^2}\!\!\E^{-\I\pi \rho^2\tanh\beta }\,\left\{\int_{{\mathbb R}^2}\E^{\I\pi \rho\primespe^2(\tanh\beta -\tanh\beta ')}\right.\nonumber \\
& &\hskip .5cm \times \;\left.\exp\!\left[2\I\pi \vec\rho '\vec\cdot\left({\vec\rho''\over \cosh\beta '}+{\vec\rho\over \cosh\beta}\right)\right]\,\D \vec\rho \primespe\,\right\}\,f(\vec\rho)\,\D \vec\rho\,.\nonumber 
\end{eqnarray}
The integral between braces, denoted $I_6(\vec \rho'')$, is a Fourier transform.

\medskip
\noindent {\its ii.} We first assume $\beta\ne \beta '$, so that $\tanh\beta-\tanh\beta '\ne 0$.
We use  Eq.\ (\ref{eq11}), once more, with
$A=1/(\tanh\beta -\tanh\beta ')$, and we obtain
\begin{equation}
\E^{\I\pi \rho'^2(\tanh\beta -\tanh\beta ')} \,\rightleftharpoons\,
  {\I\over \tanh\beta -\tanh\beta '} \,\exp\left({-\I\pi \nu^2\over \tanh\beta -\tanh\beta '}\right)\,.\label{eq103}
\end{equation}

The integral $I_6(\vec \rho '')$ is the value of the function written on the right side of Eq.\ (\ref{eq103}), taken at
$\vec \nu =(\vec \rho ''/ \cosh\beta ')+(\vec \rho /\cosh\beta)$, that is
\begin{eqnarray}
I_6(\rho '')\rap&=&\rap\int_{{\mathbb R}^2}\!\E^{\I\pi \rho\primespe^2(\tanh\beta -\tanh\beta ')}\,\exp \left[2\I\pi \vec\rho\primespe\!\vec\cdot\left({\vec \rho ''\over \cosh\beta '}+{\vec\rho \over \cosh\beta}\right)\right]\,\D \vec\rho\primespe  \\
&=&\rap
{\I\over \tanh\beta -\tanh\beta '}\exp\left({-\I\pi \over \tanh\beta -\tanh\beta '}\,\left\|{\vec\rho ''\over \cosh\beta '}+{\vec\rho\over \cosh\beta}\right\|^2\right)\,.\nonumber 
\end{eqnarray}

We use
\begin{equation}
\cosh\beta\cosh\beta '(\tanh\beta -\tanh\beta ')=
\sinh\beta \cosh\beta '-\cosh\beta \sinh\beta '
=\sinh (\beta -\beta ')\,,\end{equation}
so that
\begin{eqnarray}
{\cal K}_{\beta '}\circ{\cal K}_\beta [f](\vec\rho '') \rap&=&\rap
{-\I \E^{(\beta +\beta ')}\over \sinh(\beta -\beta ')}
\, \E^{\I\pi \rho''^2\tanh\beta '} \,\int_{{\mathbb R}^2}\E^{-\I\pi \rho^2\tanh\beta }    \label{eq106}
 \\
 & & \hskip 1cm \times \;\exp\left({-\I\pi \over \tanh\beta -\tanh\beta '}\,\left\|{\vec\rho ''\over \cosh\beta '}+{\vec\rho\over \cosh\beta}\right\|^2\right)\,f(\vec\rho )\,\D \vec\rho\,.\nonumber 
\end{eqnarray}
Equation (\ref{eq106}) takes the form
\begin{equation}
{\cal K}_{\beta '}\circ{\cal K}_\beta [f](\vec\rho '')={-\I\E^{\beta +\beta '}\over\sinh (\beta -\beta ')}
\,\E^{ \I\pi L'\rho ''^2}\int_{{\mathbb R^2}}\E^{\I\pi M' \rho^2} \E^{\I\pi N'\vec\rho\vec\cdot\vec\rho''}f(\vec\rho)\,\D\vec\rho\,,
\end{equation}
where
\begin{equation}
N'={2\over (\tanh\beta -\tanh\beta ')\cosh\beta\cosh\beta '}
={-2 \over \sinh (\beta -\beta ')}\,,
\end{equation}
\begin{eqnarray}
  M'\rap&=&\rap -\tanh\beta-{1\over (\tanh\beta -\tanh\beta ')\cosh^2\beta}
  ={-\sinh^2\beta +\cosh\beta\sinh\beta\tanh\beta '-1\over (\tanh\beta -\tanh\beta ')\cosh^2\beta }\nonumber \\
  &=&\rap
  {-\cosh\beta +\sinh\beta\tanh\beta '\over (\tanh\beta -\tanh\beta ')\cosh\beta }
  ={-1 +\tanh\beta\tanh\beta '\over \tanh\beta -\tanh\beta ' }={-1\over \tanh (\beta-\beta ')}\nonumber \\
  &=&\rap -\coth (\beta -\beta ')\,.
\end{eqnarray}
A similar derivation leads to $L'=-\coth (\beta -\beta ')$.  

Eventually, we obtain
\begin{eqnarray}
{\cal K}_{\beta '}\circ{\cal K}_\beta [f](\vec\rho'')\rap &=&\rap 
{-\I\,\E^{\beta +\beta '}\over \sinh(\beta -\beta ')}
\, \E^{-\I\pi \rho''^2 \coth(\beta - \beta ')}\nonumber \\
& &\hskip 2cm \times \;\;\int_{{\mathbb R}^2}\E^{-\I\pi \rho^2\coth (\beta - \beta ' )}\exp\left({-2\I\pi \vec\rho\vec\cdot\vec\rho ''\over \sinh(\beta -\beta ')}\right)\,f(-\vec\rho)\,\D \vec\rho \nonumber \\
\rap &=&\rap -\E^{2\beta '}\,{\cal H}_{\beta -\beta '}\bigl[\widetilde f\,\bigr](\vec\rho'') \nonumber \\
\rap &=&\rap
-\E^{2\beta '}\,{\cal H}_{\beta -\beta '}\circ{\cal P}[f](\vec\rho'')
=-\E^{2\beta '}\,{\cal P}\circ{\cal H}_{\beta -\beta '}[f](\vec\rho'')\,.
\end{eqnarray}

\smallskip
\noindent{\its iii.} For $\beta '=\beta$, Eq.\ (\ref{eq57}) becomes 
\begin{eqnarray}
  {\cal K}_{\beta }\circ{\cal K}_\beta [f](\vec\rho'')\rap &=&\rap {-\E^{2\beta }\over \cosh^2\beta}\;\int_{{\mathbb R}^2}\!\!\E^{-\I\pi (\rho^2-\rho''^2)\tanh\beta }\,f(\vec\rho)\;\left\{\int_{{\mathbb R}^2}\left[{2\I\pi \vec\rho \primespe\over \cosh\beta}\vec\cdot\left(\vec\rho''+\vec\rho\right)\right]\,\D \vec\rho \primespe\right\}\D \vec\rho\nonumber \\
  \rap &=&\rap -\E^{2\beta }\int_{{\mathbb R}^2}\!\!\E^{-\I\pi (\rho^2-\rho''^2)\tanh\beta }\,f(\vec\rho)\,\delta (\vec \rho ''+\vec \rho)\,\D\vec\rho \nonumber \\
   \rap &=&\rap -\E^{2\beta }\,f(-\vec\rho '')=-\E^{2\beta }\,\widetilde f(\vec\rho '')\nonumber \\
   \rap &=&\rap -\E^{2\beta }\,{\cal H}_0\circ{\cal P}[f](\vec\rho '')= -\E^{2\beta }\,{\cal P}\circ{\cal H}_0[f](\vec\rho '')\,.
\end{eqnarray}
(We used $\delta (\vec \rho /a)=|a|^2\delta (\vec \rho)$, where $\delta$ denotes the 2--dimensional Dirac distribution.)

The proof is complete. \qed

\begin{remark} {\rm For $\beta =0$, Proposition 4 gives ${\cal K}_0\circ{\cal K}_0 [f]=-\widetilde f$. Since ${\cal K}_0[f]=\I\,\widehat f$, we check
\begin{equation}  {\cal K}_0\circ{\cal K}_0 [f]= {\cal K}_0\bigl[\I\widehat f\;\bigr]=-\,\,\widehat{\!\!\widehat f}=-\widetilde f\,.
\end{equation}
  }
  \end{remark}
  
\subsubsection{Compatibility of eigenfunctions with the composition law}

According to Proposition 1, we have ${\cal H}_{\beta -\beta '}[g]=g$, and   ${\cal H}_{\beta -\beta '}[h]=\E^{2(\beta -\beta ')}h$.  Let us show that we obtain the same result if we apply Proposition 4.

Since $\widetilde g=g$, Lemma \ref{lem1} and Proposition 4 give
\begin{eqnarray}
  {\cal H}_{\beta -\beta '}[g]={\cal H}_{\beta -\beta '}\circ{\cal P}[g]\rap &=& \rap -\E^{-2\beta '}{\cal K}_{\beta '}\circ{\cal K}_\beta[g]\nonumber \\
  \rap &=&\rap
  -\E^{-2\beta '}{\cal K}_{\beta '}\bigl[{\cal K}_\beta[g]\bigr]\nonumber \\
  \rap &=&\rap -\E^{-2\beta '}{\cal K}_{\beta '}[h]=g\,.
\end{eqnarray}
Since $\widetilde h=h$, we also obtain
\begin{eqnarray}
 {\cal H}_{\beta -\beta '}[h]= {\cal H}_{\beta -\beta '}\circ{\cal P}[h]\rap & =&\rap -\E^{-2\beta '} {\cal K}_{\beta '}\circ{\cal K}_\beta[h]\nonumber \\  \rap &=&\rap  
 -\E^{-2\beta '}{\cal K}_{\beta '}\bigl[{\cal K}_\beta[h]\bigr]\nonumber \\
 \rap &=&\rap \E^{-2\beta '}{\cal K}_{\beta '}[\E^{2\beta}g]\nonumber \\
  \rap &=&\rap \E^{2(\beta -\beta ')}h\,.
  \end{eqnarray}

\subsection{Composition of two hyperbolic transformations of different kinds}\label{sect33}

\subsubsection{Product ${\cal K}_{\beta '}\circ{\cal H}_\beta$}

\begin{proposition}\label{prop5} For every $\beta$ and $\beta '$ belonging to ${\mathbb R}$, we have
  ${\cal K}_{\beta '}\circ{\cal H}_\beta ={\cal K}_{\beta '+\beta}$.
\end{proposition}

\noindent {\its Proof.} Since ${\cal H}_0={\cal I}$, when $\beta =0$, we have  ${\cal K}_{\beta '}\circ{\cal H}_0= {\cal K}_{\beta '}$, and the proposition holds.

Next, we assume $\beta \ne 0$. We derive
\begin{eqnarray}
  {\cal K}_{\beta '}\circ{\cal H}_\beta [f](\rho '')\rap &=&\rap{\cal K}_{\beta '}\bigl[{\cal H}_\beta [f]\bigr](\vec \rho'')
  \label{eq158}\label{eqDn32} \\
&=&\rap
{\I\,\E^{\beta '}\over \cosh\beta '}\,\E^{\I\pi \rho''^2\tanh\beta '}\nonumber \\
& & \hskip .7cm \times\;\;\int_{{\mathbb R}^2}\E^{-\I\pi \rho'^2\tanh\beta '}\,\exp\left({2\I\pi \vec\rho \primespe\vec\cdot\vec \rho''\over \cosh\beta '}\right)\,{\cal H}_\beta [f](\vec\rho\primespe)\,\D \vec\rho\primespe\nonumber \\
&=&\rap {-\E^{\beta +\beta '}\over \sinh\beta \cosh\beta '}\,\E^{\I\pi \rho''^2\tanh\beta '}\nonumber \\
& &\hskip .7cm \times \;\;\int_{{\mathbb R}^2}\!\!\E^{-\I\pi \rho'^2\tanh\beta '}\,\exp\left({2\I\pi \vec\rho\primespe\vec\cdot\vec\rho''\over \cosh\beta '}\right)\, \E^{-\I\pi \rho'^2\coth\beta}\nonumber \\
& &\hskip .7cm \times \;\;\left\{\int_{{\mathbb R}^2}\E^{-\I\pi \rho^2\coth\beta}\,\exp\left({2\I\pi \vec\rho\vec\cdot\vec\rho \primespe\over \sinh\beta}\right)\,f(\vec\rho)\,\D \vec\rho\,\right\}\,\D \vec\rho \primespe\nonumber \\
&=&\rap {-\E^{\beta +\beta '}\over \sinh\beta \cosh\beta '}\,\E^{\I\pi \rho''^2\tanh\beta '}\nonumber \\
& &\hskip .7cm \times \;\;\int_{{\mathbb R}^2}\!\!\E^{-\I\pi \rho^2\coth\beta }\,\left\{\int_{{\mathbb R}^2}\E^{-\I\pi \rho'^2(\tanh\beta ' +\coth\beta )}\right.\nonumber \\
& &\hskip .7cm \times \;\; \,\left.\exp\left[2\I\pi \vec\rho \primespe\vec\cdot\left({\vec \rho''\over \cosh\beta '}+{\vec\rho\over \sinh\beta}\right)\right]\D \vec\rho\primespe\right\}\,f(\vec\rho)\,\D \vec\rho\,.\nonumber 
\end{eqnarray}

The integral between braces, denoted $I_7(\vec\rho'')$, is a Fourier transform. We use  Eq.\ (\ref{eq11}) with
$A=-1/(\tanh\beta '+\coth\beta )$ and obtain
\begin{eqnarray}
\E^{-\I\pi \rho'^2(\tanh\beta '+\coth\beta )} &\rightleftharpoons&
  {-\I \over \tanh\beta '+\coth\beta } \,\exp\left({\I\pi \nu ^2\over \tanh\beta '+\coth\beta }\right)\,.
 \label{eq159}
\end{eqnarray}
The integral $I_7(\vec\rho '')$ is the value of the function on the right side of Equation (\ref{eq159}), taken at
$\vec \nu =(\vec\rho ''/ \cosh\beta ')+(\vec\rho /\sinh\beta)$, that is
\begin{eqnarray}
I_7(\vec\rho '')\rap &=&\rap\int_{{\mathbb R}^2}\!\!\E^{-\I\pi \rho'^2(\coth\beta +\tanh\beta ')}\,\exp \left[2\I\pi \vec\rho\primespe\vec\cdot \left({\vec\rho''\over \cosh\beta '}+{\vec\rho\over \sinh\beta}\right)\right]\,\D \vec\rho \primespe  \nonumber \\
&=&\rap
{-\I \over \tanh\beta '+\coth\beta}\exp \left({\I\pi \over \tanh\beta '+\coth\beta }\left\|{\vec\rho ''\over \cosh\beta '}+{\vec\rho\over \sinh\beta}\right\|^2\right).
\end{eqnarray}

We have
\begin{equation}
\sinh\beta\cosh\beta '(\tanh\beta '+\coth\beta )=
\sinh\beta \sinh\beta '+\cosh\beta '\cosh\beta =
\cosh (\beta +\beta ')\,,\end{equation}
so that
\begin{eqnarray}
{\cal K}_{\beta '}\circ{\cal H}_\beta [f](\vec\rho '')\rap&=&\rap
{ \I\,\E^{\beta +\beta '}\over \cosh(\beta +\beta ')}
\, \E^{\I\pi \rho''^2\tanh\beta '}\int_{{\mathbb R}^2}\E^{-\I\pi \rho ^2\coth\beta }  
\nonumber \\
& & \hskip -.2cm \times \;\exp\!\left({\I\pi \over \tanh\beta '+\coth\beta }\left\|{\vec\rho ''\over \cosh\beta '}+{\vec \rho\over \sinh\beta}\right\|^2\right)\,f(\vec\rho)\,\D \vec\rho\nonumber \\
\rap &=& \rap { \I\,\E^{\beta +\beta '}\over \cosh(\beta +\beta ')}
\,\E^{\I\pi L''\rho''^2}\int_{{\mathbb R}^2}\E^{\I\pi M'' \rho^2} \E^{\I\pi N''\vec\rho\vec\cdot\vec\rho ''}f(\vec\rho)\,\D \vec\rho\,, \label{eqDn36}
\end{eqnarray}
where
\begin{equation}
  N''={2\over (\tanh\beta '+\coth\beta )\sinh\beta\cosh\beta '}
={2 \over \cosh (\beta +\beta ' )}\,,
\end{equation}
\begin{eqnarray}
  M''\rap &=&\rap -\coth\beta+{1\over (\tanh\beta '+\coth\beta )\sinh^2\beta}
  ={-\cosh^2\beta -\cosh\beta\sinh\beta\tanh\beta '+1\over (\tanh\beta '+\coth\beta )\sinh^2\beta }\nonumber \\
&=&\rap {-\sinh^2\beta -\cosh\beta\sinh\beta\tanh\beta '\over (\tanh\beta '+\coth\beta )\sinh^2\beta }
  ={-1 -\coth\beta\tanh\beta '\over \tanh\beta '+\coth\beta  }\nonumber \\
  \rap &=&\rap -{\tanh\beta +\tanh\beta '\over \tanh \beta '\tanh\beta +1}= -\tanh (\beta +\beta ')\,,
\end{eqnarray}
\begin{eqnarray}
  L''\rap &=&\rap \tanh\beta '+{1\over (\tanh\beta '+\coth\beta )\cosh^2\beta '}
  ={\sinh^2\beta +\coth\beta\sinh\beta \cosh\beta '+1\over (\tanh\beta '+\coth\beta )\cosh^2\beta '}\nonumber \\
&=&\rap{\cosh^2\beta +\coth\beta\sinh\beta '\cosh\beta '\over (\tanh\beta '+\coth\beta )\cosh^2\beta '}
  ={1 +\coth\beta\tanh\beta '\over \tanh\beta '+\coth\beta  }\nonumber \\
  &=&\rap {\tanh\beta +\tanh\beta '\over \tanh \beta '\tan\beta +1}= \tanh (\beta +\beta ')\,.
\end{eqnarray}
We eventually obtain
\begin{eqnarray}
{\cal K}_{\beta '}\circ{\cal H}_\beta [f](\vec\rho '')\rap &=&\rap   {\I \E^{\beta +\beta '}\over \cosh(\beta +\beta ')}\,\E^{\I\pi \rho''^2\tanh (\beta +\beta ')} \nonumber  \\
& &\hskip .4cm \times \;\;
\int_{\mathbb R}\E^{-\I\pi \rho^2\tanh (\beta +\beta ')} \exp\!\left({2\I\pi \vec\rho\vec\cdot\vec\rho ''\over \cosh (\beta +\beta ')}\right)f(\vec\rho)\,\D \vec\rho \nonumber \\
&=& \rap {\cal K}_{\beta+\beta '}[f](\vec\rho'')\,. 
\end{eqnarray}
The proof is complete. \qed

\subsubsection{Product ${\cal H}_{\beta '}\circ{\cal K}_\beta$}

\begin{proposition}\label{prop6} For every $\beta$ and every $\beta '$ belonging to ${\mathbb R}$, we have
  ${\cal H}_{\beta '}\circ{\cal K}_\beta =\E^{2\beta '}{\cal K}_{\beta -\beta '}$.
  \end{proposition}
  
\noindent {\its Proof.} If $\beta '=0$,  we obtain ${\cal H}_0\circ{\cal K}_\beta={\cal K}_\beta$, because  ${\cal H}_0={\cal I}$. The proposition holds.

If $\beta '\ne 0$, we derive
\begin{eqnarray}
  {\cal H}_{\beta '}\circ{\cal K}_\beta [f](\vec \rho '')\rap &=&\rap{\cal H}_{\beta '}\bigl[{\cal K}_\beta [f]\bigr](\vec \rho'')
  \nonumber  \\
&=&\rap
{\I\,\E^{\beta '}\over \sinh\beta '}\,\E^{-\I\pi \rho''^2\coth\beta '}\nonumber \\
& & \hskip .7cm \times\;\;\int_{{\mathbb R}^2}\E^{-\I\pi \rho'^2\coth\beta '}\,\exp\left({2\I\pi \vec\rho \primespe\vec\cdot\vec \rho''\over \sinh\beta '}\right)\,{\cal K}_\beta [f](\vec\rho\primespe)\,\D \vec\rho\primespe\nonumber \\
&=&\rap {-\E^{\beta +\beta '}\over \sinh\beta '\cosh\beta }\,\E^{-\I\pi \rho''^2\coth\beta '}\nonumber \\
& &\hskip .7cm \times \;\;\int_{{\mathbb R}^2}\!\!\E^{-\I\pi \rho'^2\coth\beta '}\,\exp\left({2\I\pi \vec\rho\primespe\vec\cdot\vec\rho''\over \sinh\beta '}\right)\, \E^{\I\pi \rho'^2\tanh\beta}\nonumber \\
& &\hskip .7cm \times \;\;\left\{\int_{{\mathbb R}^2}\E^{-\I\pi \rho^2\tanh\beta}\,\exp\left({2\I\pi \vec\rho\vec\cdot\vec\rho \primespe\over \cosh\beta}\right)\,f(\vec\rho)\,\D \vec\rho\,\right\}\,\D \vec\rho \primespe\nonumber \\
&=&\rap {-\E^{\beta +\beta '}\over \sinh\beta ' \cosh\beta}\,\E^{-\I\pi \rho''^2\coth\beta '} 
\int_{{\mathbb R}^2}\!\!\E^{-\I\pi \rho^2\tanh\beta }\,\left\{\int_{{\mathbb R}^2}\E^{\I\pi \rho'^2(\tanh\beta  -\coth\beta ' )}\right.\nonumber \\
& &\hskip .7cm \times \;\; \,\left.\exp\left[2\I\pi \vec\rho \primespe\vec\cdot\left({\vec \rho''\over \sinh\beta ' }+{\vec\rho\over \cosh\beta }\right)\right]\D \vec\rho\primespe\right\}\,f(\vec\rho)\,\D \vec\rho\,.\label{eq158b}
\end{eqnarray}

The integral between braces, denoted $I_8(\vec\rho'')$, is a Fourier transform. We use  Eq.\ (\ref{eq11}) with
$A=1/(\tanh\beta -\coth\beta ')$ and obtain
\begin{eqnarray}
\E^{\I\pi \rho'^2(\tanh\beta -\coth\beta ')} &\rightleftharpoons&
  {\I \over \tanh\beta -\coth\beta '} \,\exp\left({-\I\pi \nu ^2\over \tanh\beta -\coth\beta '}\right)\,.
 \label{eq159b}
\end{eqnarray}
The integral $I_8(\vec\rho '')$ is the value of the function on the right side of Equation (\ref{eq159b}), taken at
$\vec \nu =(\vec\rho ''/ \sinh\beta ')+(\vec\rho /\cosh\beta)$, that is
\begin{eqnarray}
I_8(\vec\rho '')\rap &=&\rap\int_{{\mathbb R}^2}\!\!\E^{\I\pi \rho'^2(\tanh\beta -\coth\beta ')}\,\exp \left[2\I\pi \vec\rho \primespe\vec\cdot \left({\vec\rho''\over \sinh\beta '}+{\vec\rho\over \cosh\beta}\right)\right]\,\D \vec\rho \primespe  \nonumber \\
&=&\rap
{\I \over \tanh\beta -\coth\beta '}\exp \left({-\I\pi \over \tanh\beta -\coth\beta ' }\left\|{\vec\rho ''\over \sinh\beta '}+{\vec\rho\over \cosh\beta}\right\|^2\right).
\end{eqnarray}
We have
\begin{equation}
(\tanh\beta -\coth\beta ')\sinh\beta'\cosh\beta =
\sinh\beta '\sinh\beta -\cosh\beta '\cosh\beta =
-\cosh (\beta -\beta ')\,,\end{equation}
so that
\begin{eqnarray}
{\cal H}_{\beta '}\circ{\cal K}_\beta [f](\vec\rho '')\rap&=&\rap
{ \I\,\E^{\beta +\beta '}\over \cosh(\beta -\beta ')}
\, \E^{-\I\pi \rho''^2\coth\beta '}\int_{{\mathbb R}^2}\E^{-\I\pi \rho ^2\tanh\beta }   
\nonumber \\
& & \hskip -.2cm \times \;\exp\!\left({-\I\pi \over \tanh\beta -\coth\beta '}\left\|{\vec\rho ''\over \sinh\beta '}+{\vec \rho\over \cosh\beta}\right\|^2\right)\,f(\vec\rho)\,\D \vec\rho\nonumber \\
\rap &=& \rap { \I\,\E^{\beta +\beta '}\over \cosh(\beta -\beta ')}
\,\E^{\I\pi L'''\rho''^2}\int_{{\mathbb R}^2}\E^{\I\pi M''' \rho^2} \E^{\I\pi N'''\vec\rho\vec\cdot\vec\rho ''}f(\vec\rho)\,\D \vec\rho\,, \label{eq70}
\end{eqnarray}
where
\begin{equation}
  N'''={-2\over (\tanh\beta -\coth\beta ')\sinh\beta '\cosh\beta }={2 \over \cosh (\beta -\beta ' )}\,,
\end{equation}
\begin{eqnarray}
  M'''\rap &=&\rap -\tanh\beta-{1\over (\tanh\beta -\coth\beta ')\cosh^2\beta}
={-\sinh^2\beta +\cosh\beta\sinh\beta\coth\beta '-1\over (\tanh\beta -\coth\beta ' )\cosh^2\beta }\nonumber \\
&=&\rap {-\cosh^2\beta +\cosh\beta\sinh\beta\coth\beta '\over (\tanh\beta -\coth\beta ' )\cosh^2\beta }
  ={-1 +\tanh\beta\coth\beta '\over \tanh\beta -\coth\beta  ' }\nonumber \\
  &=&\rap {\tanh\beta -\tanh\beta ' \over \tanh \beta \tanh\beta '-1}=-\tanh (\beta -\beta ')\,,
\end{eqnarray}
\begin{eqnarray}
  L'''\rap &=&\rap -\coth\beta '-{1\over (\tanh\beta -\coth\beta ')\sinh^2\beta '}
={\cosh^2\beta '-\tanh\beta\sinh\beta ' \cosh\beta '-1\over (\tanh\beta -\coth\beta ')\sinh^2\beta '}\nonumber \\
&=&\rap {\sinh^2\beta '-\tanh\beta\sinh\beta '\cosh\beta '\over (\tanh\beta -\coth\beta ')\sinh^2\beta '}
={1 -\tanh\beta\coth\beta '\over \tanh\beta -\coth\beta '}\nonumber \\
  &=&\rap {\tanh\beta ' -\tanh\beta \over \tanh \beta '\tanh\beta -1}=\tanh (\beta -\beta ')\,.
\end{eqnarray}
We obtain
\begin{eqnarray}
{\cal H}_{\beta '}\circ{\cal K}_\beta [f](\vec\rho '')\rap &=&\rap   {\I \,\E^{\beta +\beta '}\over \cosh(\beta -\beta ')}\,\E^{\I\pi \rho''^2\tanh (\beta -\beta ')} \nonumber  \\
& &\hskip .4cm \times \;\;
\int_{\mathbb R}\E^{-\I\pi \rho^2\tanh (\beta -\beta ')} \exp\!\left({2\I\pi \vec\rho\vec\cdot\vec\rho ''\over \cosh (\beta -\beta ')}\right)f(\vec\rho)\,\D \vec\rho \nonumber \\
\rap &=& \E^{2\beta '}{\cal K}_{\beta-\beta '}[f](\vec\rho'')\,.
\end{eqnarray}
The proof is complete. \qed


\begin{remark} {\rm From Propositions 5 and 6, we deduce that the product ${\cal H}_{\beta '}\circ{\cal K}_\beta$ is not commutative.}
  \end{remark}

\subsection{Algebra of hyperbolic fractional-order Fourier transformations}\label{sect34}

Hyperbolic fractional-order Fourier transformations obey the following rules:
\begin{enumerate}
\item[i.] ${\cal H}_0= {\cal I}$ (identity operator) ;
\item[ii.]${\cal H}_{\beta '}\circ{\cal H}_\beta ={\cal H}_{\beta '+\beta}={\cal H}_{\beta }\circ{\cal H}_{\beta '}$ ;
   \item[iii.]
     ${\cal H}_\beta^{-1}={\cal H}_{-\beta}$ ;
   \item [iv.]  ${\cal K}_{0}=\I\,{\cal F}$ (${\cal F}$ denotes the standard Fourier transformation);
      \item [v.] ${\cal K}_{0}\circ {\cal K}_{0}=-{\cal F}^2=-{\cal P}$ (${\cal P}$ denotes the parity operator) ; ${{\cal K}_0}^3=-\I\, {\cal F}^{-1}$ ; ${{\cal K}_0}^4={\cal I}$ ;
      \item[vi.] ${\cal K}_{\beta '}\circ{\cal K}_\beta=-\E^{2\beta '}{\cal H}_{\beta -\beta '}\circ{\cal P}$ ;
        \item [vii.]  ${\cal K}_{\beta}\circ {\cal K}_{\beta}=-\E^{2\beta }{\cal P}$ ;
\item[viii.] ${\cal K}_{\beta '}\circ{\cal H}_\beta={\cal K}_{\beta '+\beta }$ ;
           \item [ix.] ${\cal H}_{\beta '}\circ{\cal K}_\beta=\E^{2\beta '}{\cal K}_{\beta -\beta '}$ ;
 \item [x.] ${\cal K}_{0}= {\cal K}_{\beta}\circ {\cal H}_{-\beta}$ ;
      \item [xi.] ${\cal K}_{\beta}= {\cal K}_{0}\circ {\cal H}_{\beta}$.
\end{enumerate}


\section{Expressing diffraction by a hyperbolic fractional-order\\ Fourier transformation} \label{sect4}
In the introduction, we mention that for $J>0$, Eq.\ (\ref{eq1}) takes the form of a (circular) fractional-order Fourier transform if appropriate reduced variables and functions are chosen \cite{PPF1}. In this section, we show that for $J<0$, Eq.\ (\ref{eq1}) takes the form of a hyperbolic fractional-order Fourier transform. 

\subsection{The case $J<-1$}\label{sect41}

Figure \ref{fig1} represents a diffraction-propagation phenomenon from a spherical cap ${\cal A}$ (the emitter) to a spherical cap ${\cal B}$ (the receiver) at a distance $D$ (taken from vertex to vertex). 
In the framework of a scalar theory, the field amplitudes $U_A$ on ${\cal A}$ and $U_B$ on ${\cal B}$ are connected by Eq.\ (\ref{eq1}) \cite{PPF1}.
\begin{figure}[h]
  \centering
    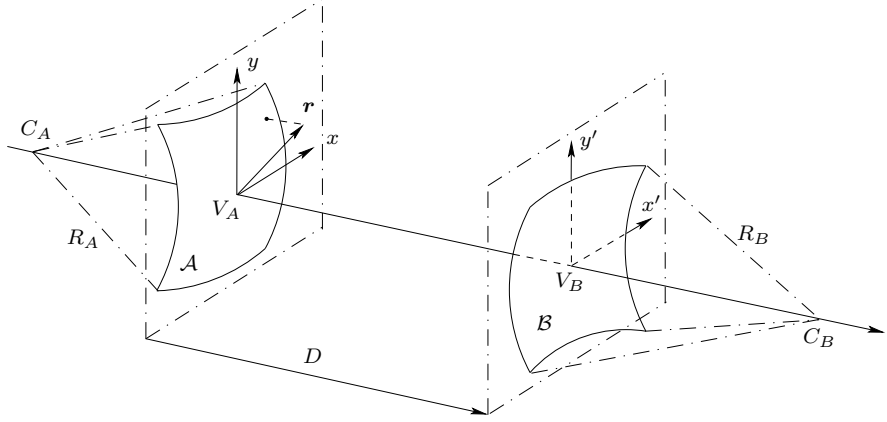
    \caption{\small Elements for representing the diffraction from a spherical emitter ${\cal A}$ to a spherical receiver ${\cal B}$ at a distance $D$. Light propagates from left to right.\label{fig1}}
\end{figure}

The problem is to write Eq.\ (\ref{eq1}) as a hyperbolic fractional-order Fourier transform.
In this section, we assume  $J<-1$, that is, according to Eq.\ (\ref{eq3})
\begin{equation} J={(R_A-D)(R_B+D)\over D(D-R_A+R_B)}<-1\,.\label{eq25}
\end{equation}
(Consequences of Eq.\ (\ref{eq25}) are: $D\ne R_A$ and  $D\ne -R_B$. We also assume $D\ne 0$. The case $D=0$ may be studied as a limit case.)

\begin{lemma}\label{lem2}  Under the assumption of Eq.\ (\ref{eq25}),  $R_A(R_A-D)$ and $R_B(R_B+D)$ have the same sign.
\end{lemma}

\noindent{\its Proof.} From the identity $D(D-R_A+R_B)=R_AR_B-(R_A-D)(R_B+D)$, we deduce
\begin{equation}
  {1\over J}={D(D-R_A+R_B)\over (R_A-D)(R_B+D)}={R_AR_B\over (R_A-D)(R_B+D)}-1\,,\end{equation}
that is
\begin{equation}
  {R_AR_B\over (R_A-D)(R_B+D)}=1+{1\over J}>0\,,\label{eq26}\end{equation}
because $J<-1$. Then  $R_A(R_A-D)$ and $R_B(R_B+D)$ have the same sign. \qed

\subsubsection*{Definitions}

We introduce the following parameters, variables and functions:
\begin{enumerate}
\item[$\bullet$]  The sign of $R_A(R_A-D)$, denoted $\frak{s}$, such that
  \begin{enumerate}
\item[$\ast$] $\frak{s}=1$, if  $R_A(R_A-D)>0$ ;
\item[$\ast$] $\frak{s}=0$, if  $R_A(R_A-D)=0$ (not possible here, because $0\ne R_A\ne D$);
\item[$\ast$] $\frak{s}=-1$, if  $R_A(R_A-D)<0$.
  \end{enumerate}
  \item[$\bullet$] The order $\beta$ ($\beta\in{\mathbb R}$), defined by
  \begin{equation}
    \coth^2\beta =-J=-{(R_A-D)(R_B+D)\over D(D-R_A+R_B)}\,,\hskip 1cm \beta D>0\,.\end{equation}
\item[$\bullet$] Auxiliary parameters $\varepsilon_A$ and $\varepsilon_B$, such that
  \begin{equation}
    \varepsilon_A=\frak{s}{D\over R_A-D}\,\coth\beta \,,\hskip 1cm   \varepsilon_B=\frak{s}{D\over R_B+D}\,\coth\beta\,.\end{equation}
 \item[$\bullet$] 
  Reduced variables (or coordinates) on ${\cal A}$ and ${\cal B}$, respectively
\begin{equation}
  \vec\rho ={\vec r\over\sqrt{\lambda \varepsilon_AR_A}}\,,\hskip 1cm  \vec\rho '={\vec r'\over\sqrt{\lambda \varepsilon_BR_B}}\,.\label{eq100}
\end{equation}
\item[$\bullet$]
Reduced field amplitudes on ${\cal A}$ and ${\cal B}$
\begin{equation}
  u_A(\vec \rho)=\sqrt{\lambda\varepsilon_AR_A}\,U_A\left(\sqrt{\lambda \varepsilon_AR_A}\,\vec \rho\right)\,,\hskip .3cm u_B(\vec \rho ')=\sqrt{\lambda\varepsilon_BR_B}\,U_B\left(\sqrt{\lambda \varepsilon_BR_B}\,\vec\rho '\right)\,.\label{eq101}\end{equation}
\end{enumerate}

Equations  (\ref{eq100}) and (\ref{eq101}) above make sense if the quantities under the square roots are positive. The following lemma shows that this is the case.

\begin{lemma}\label{lem3} For $J<-1$, and under the previous definitions and assumptions, we have: $\varepsilon_AR_A>0$ and $\varepsilon_BR_B>0$.
  \end{lemma}

\noindent {\its Proof.} Since $\beta D>0$, we have $D\coth\beta >0$, and from
\begin{equation}
  \varepsilon_AR_A=\frak{s}{R_AD\over R_A-D}\,\coth\beta\,,\end{equation}
we conclude that $\varepsilon_AR_A$ has the sign of $\frak{s}R_A(R_A-D)$, which is positive by definition of $\frak{s}$.

For the same reason the sign of $\varepsilon_BR_B$ is the sign of $\frak{s}R_B(R_B+D)$, which is also the sign of $\frak{s}R_A(R_A-D)$, according to Lemma \ref{lem2}, and is positive. \qed

\begin{theorem}\label{th1} Let ${\cal A}$ (vertex $V_A$, radius $R_A$) be an emitter and ${\cal B}$ (vertex $V_B$, radius $R_B$) be a receiver at a distance $D=\overline{V_AV_B}$.
  Under the assumption $J<-1$,  and under the previous definitions, the field transfer from ${\cal A}$ to ${\cal B}$, which is expressed by Eq.\ (\ref{eq1}), can then be expressed, with reduced field amplitudes $u_A$ and $u_B$, in the form
  \begin{equation}
    u_B(\vec \rho \primespe )=\E^{-\beta}{\cal H}_{\beta}[u_A](\vec\rho  \primespe)\,,\hskip 1.1cm \mbox{if}\;\;\;\frak{s}=1\,,\label{eq95a}\end{equation}
  or
  \begin{equation}
    u_B(\vec \rho \primespe )=-\E^{\beta}{\cal P}\circ{\cal H}_{-\beta}[u_A](\vec\rho  \primespe)
    \,,\hskip .3cm \mbox{if}\;\;\;\frak{s}=-1\,,\label{eq95b}\end{equation}
    where ${\cal H}_\beta$ denotes the hyperbolic fractional Fourier transformation of the first kind and order
$\beta$.
\end{theorem}

\begin{remark}\label{rem7}{\rm Equations (\ref{eq95a}) and (\ref{eq95b})  are synthetized in
  \begin{equation}
  u_B(\vec \rho \primespe )=\frak{s}\,\E^{-\frak{s}\beta}{\cal H}_{\frak{s}\beta}[u_A](\frak{s}\vec\rho  \primespe)\,.\label{eq95}\end{equation}}
  \end{remark}

\noindent{\its Proof of Theorem \ref{th1}.} We have to change spatial variables $\vec r$ and $\vec r'$ in Eq.\ (\ref{eq1}) to reduced variables $\vec\rho$ and $\vec \rho \primespe$.

\smallskip
\noindent{\its i. Derivation of $\D\vec r/\lambda D$.}
Since $\vec r$ and $\vec \rho$ are two-dimensional variables (if $\D\vec r=\D x\,\D y$, then $\D (a\vec r)=|a|^2\D x\,\D y$), we have $\D\vec r=(\sqrt{\lambda \varepsilon_AR_A})^2\,\D\vec \rho$, so that
\begin{equation}
  {\D\vec r\over \lambda D}
  ={\varepsilon_AR_A\over D}\,\D\vec\rho\,.
\end{equation}
Then
\begin{eqnarray}
  \left({\varepsilon_AR_A\over D}\right)^2={{R_A}^{\! 2}\over (R_A-D)^2}\coth^2\beta
  \rap &=&\rap -{{R_A}^{\! 2}\over R_A-D}\,{R_B+D\over D(D-R_A+R_B)}\nonumber \\
  \rap &=&\rap -{R_A\over R_B}\;{R_B+D\over R_A-D}\;{R_AR_B\over D(D-R_A+R_B)}\nonumber \\
  \rap &=&\rap -{\varepsilon_AR_A\over \varepsilon_BR_B}\;{R_AR_B\over D(D-R_A+R_B)}\,.
\end{eqnarray}
From
\begin{equation}
  {\cosh^2\beta\over \sinh^2\beta}=\coth^2\beta ={-(R_A-D)(R_B+D)\over D(D-R_A+R_B)}\,,\end{equation}
we deduce
\begin{equation}
  {1\over \sinh^2\beta}={-(R_A-D)(R_B+D) -D(D-R_A+R_B)
    \over  D(D-R_A+R_B)}=-{R_AR_B\over  D(D-R_A+R_B)}\,,
\end{equation}
so that
\begin{equation}
  \left({\varepsilon_AR_A\over D}\right)^2= {1\over \sinh^2\beta}\;{\varepsilon_AR_A\over \varepsilon_BR_B}\,.
\end{equation}
The sign of $\sinh\beta$ is the sign of $D$, because  $\beta D >0$, and  since $\varepsilon_AR_A>0$ and $\varepsilon_BR_B>0$, we obtain
\begin{equation}
  {\D\vec r\over \lambda D}={\varepsilon_AR_A\over D}\,\D\vec\rho={1\over \sinh\beta}\,\sqrt{{\varepsilon_AR_A\over \varepsilon_BR_B}}\,\D\vec\rho\,.
  \end{equation}

\medskip
\noindent{\its ii. Derivation of $\vec r\vec\cdot\vec r'/\lambda D$.}
We have
\begin{equation}
  {\vec r\vec\cdot\vec r'\over\lambda D} = {1\over D}\sqrt{\varepsilon_AR_A\varepsilon_BR_B}\, \vec \rho\vec\cdot\vec \rho \primespe = {1\over D}\sqrt{{D^2R_AR_B\over (R_A-D)(R_B+D)}\coth^2\beta} \, \vec \rho\vec\cdot\vec \rho \primespe\,.\end{equation}
We use Eq.\ (\ref{eq26}) and, since $\coth^2\beta =-J$, we obtain
\begin{eqnarray}
  {\vec r\vec\cdot\vec r'\over\lambda D} ={1\over D}\sqrt{D^2\coth^2\beta \left(1-{1\over \coth^2\beta}\right)}\, \vec \rho\vec\cdot\vec \rho \primespe
  \rap &=&\rap {1\over D}\sqrt{D^2(\coth^2\beta -1)}\,\vec \rho\vec\cdot\vec \rho \primespe  \nonumber \\
  \rap &=&\rap {1\over D}\sqrt{{D^2\over \sinh^2\beta}}\,\vec \rho\vec\cdot\vec \rho \primespe\,.
\end{eqnarray}
Since $\beta D>0$ (and then $D\sinh\beta >0$), we finally obtain
\begin{equation}
   {\vec r\vec\cdot\vec r'\over\lambda D}={\vec \rho\vec\cdot\vec \rho \primespe \over\sinh\beta} \,.\end{equation}

\medskip
\noindent{\its iii. Derivation of quadratic-phase factors.}
We begin with
\begin{equation}
  {1\over \lambda}\left({1\over D}-{1\over R_A}\right)r^2={1\over \lambda}{R_A-D\over R_AD}\lambda\varepsilon_AR_A\rho^2
  =\frak{s}\rho^2\coth\beta\,.
\end{equation}
For the other factor, we have
\begin{equation}
  {1\over \lambda}\left({1\over R_B}+{1\over D}\right)r'^{\,2}={1\over \lambda}{R_B+D\over R_BD}\lambda\varepsilon_BR_B\rho \primespe^{\,2}
  =\frak{s}\rho \primespe^{\, 2}\coth\beta\,.
\end{equation}

\medskip
\noindent{\its iv. Integral.}
Eventually, Eq.\ (\ref{eq1}) is written
\begin{eqnarray}
  U_B\left(\sqrt{\lambda \varepsilon_BR_B}\vec \rho  \primespe\right)\rap &=&\rap {\I\over \sinh\beta}\,\sqrt{{\varepsilon_AR_A\over \varepsilon_BR_B}}
  \E^{-\I{\frak s}\pi\rho'^2\coth\beta} \\
  & &\hskip .5cm \times \;\;\int_{{\mathbb R}^2}  \E^{-\I{\frak s}\pi\rho^2\coth\beta}
  \,\exp\left({2\I\pi\over\sinh\beta}\,\vec\rho\vec\cdot\vec\rho  \primespe\right)\,U_A \left(\sqrt{\lambda \varepsilon_BR_B}\vec \rho \right)\,\D\vec\rho\,, \nonumber
  \end{eqnarray}
that is
\begin{equation}
  u_B(\vec \rho  \primespe)={\I\over \sinh\beta}
   \E^{-\I{\frak s}\pi\rho'^2\coth\beta}
   \,\int_{{\mathbb R}^2}  \E^{-\I{\frak s}\pi\rho^2\coth\beta}\exp\left({2\I\pi\over\sinh\beta}\,\vec\rho\vec\cdot\vec\rho '\right)\,u_A(\vec \rho )\,\D\vec\rho\,.
\end{equation}
If $\frak{s}=1$, we obtain
\begin{eqnarray}
  u_B(\vec \rho  \primespe)\rap &=&\rap {\I\over \sinh\beta}
   \E^{-\I\pi\rho'^2\coth\beta}
   \,\int_{{\mathbb R}^2}  \E^{-\I\pi\rho^2\coth\beta}\exp\left({2\I\pi\over\sinh\beta}\,\vec\rho\vec\cdot\vec\rho '\right)\,u_A(\vec \rho )\,\D\vec\rho\nonumber \\
   \rap &=&\rap \E^{-\beta} {\cal H}_\beta[u_A](\vec\rho  \primespe)\,.\label{eq46}
\end{eqnarray}
If $\frak{s}=-1$, we obtain
\begin{eqnarray}
  u_B(\vec \rho  \primespe)\rap &=&\rap {\I\over \sinh\beta}
   \E^{\I\pi\rho'^2\coth\beta}
   \,\int_{{\mathbb R}^2}  \E^{\I\pi\rho^2\coth\beta}\exp\left({2\I\pi\over\sinh\beta}\,\vec\rho\vec\cdot\vec\rho  \primespe\right)\,u_A(\vec \rho )\,\D\vec\rho\nonumber \\
   \rap &=&\rap -\E^{\beta} {\cal H}_{-\beta}[u_A](-\vec\rho  \primespe)
   \nonumber \\
   \rap &=&\rap  -\E^{\beta}\,{\cal P}\circ {\cal H}_{-\beta}[u_A](\vec\rho  \primespe)\,.\label{eq47}
\end{eqnarray}
The proof is complete.\qed

\subsection{The case $-1<J<0$}\label{sect42}
We consider an emitter ${\cal A}$ and a receiver ${\cal B}$ at a distance $D$ (Fig.\ \ref{fig1}). In this section, we assume $-1<J<0$.

\begin{lemma}\label{lem4} Under the assumption $-1<J<0$, the signs of  $R_A(R_A-D)$ and $R_B(R_B+D)$ are opposite.
  \end{lemma}

\noindent{\its Proof.} From $D(D-R_A+R_B)=R_AR_B-(R_A-D)(R_B+D)$, we deduce
\begin{equation}
  {1\over J}={D(D-R_A+R_B)\over (R_A-D)(R_B+D)}={R_AR_B\over (R_A-D)(R_B+D)}-1\,,\end{equation}
that is
\begin{equation}
  {R_AR_B\over (R_A-D)(R_B+D)}=1+{1\over J}<0\,,\label{eq116}\end{equation}
because $-1<J<0$ and $1/J<-1$. Then  $R_A(R_A-D)$ and $R_B(R_B+D)$ have opposite signs. \qed

\subsubsection*{Definitions}

We introduce:
\begin{enumerate}
\item[$\bullet$]  The sign of $R_A(R_A-D)$, denoted by $\frak{s}$. 
  \item[$\bullet$] The order $\beta$ ($\beta\in{\mathbb R}$), defined by
  \begin{equation}
    \coth^2\beta =-{1\over J}=-{D(D-R_A+R_B)\over(R_A-D)(R_B+D)}\,,\hskip 1cm \beta D>0\,.\end{equation}
\item[$\bullet$] Auxiliary parameters $\varepsilon_A$ and $\varepsilon_B$ 
  \begin{equation}
    \varepsilon_A=\frak{s}{D\over R_A-D}\,{1\over \coth\beta} \,,\hskip 1cm   \varepsilon_B=-\frak{s}{D\over R_B+D}\,{1\over \coth\beta}\,.\label{eq119}\end{equation}
 \item[$\bullet$] 
  Reduced variables (or coordinates) on ${\cal A}$ and ${\cal B}$, respectively
\begin{equation}
  \vec\rho ={\vec r\over\sqrt{\lambda \varepsilon_AR_A}}\,,\hskip 1cm  \vec\rho '={\vec r'\over\sqrt{\lambda \varepsilon_BR_B}}\,.
\label{eq120}\end{equation}
\item[$\bullet$]
Reduced field amplitudes on ${\cal A}$ and ${\cal B}$ 
\begin{equation}
  u_A(\vec \rho)=\sqrt{\lambda\varepsilon_AR_A}\,U_A\left(\sqrt{\lambda \varepsilon_AR_A}\,\vec \rho\right)\,,\hskip .3cm u_B(\vec \rho ')=\sqrt{\lambda\varepsilon_BR_B}\,U_B\left(\sqrt{\lambda \varepsilon_BR_B}\,\vec\rho '\right)\,.\end{equation}
\end{enumerate}

\begin{lemma}\label{lem5} For $-1<J<0$, and under the previous definitions and assumptions, we have: $\varepsilon_AR_A>0$ and $\varepsilon_BR_B>0$.
\end{lemma}

\noindent {\its Proof.} Since $\beta D>0$, we have $D\coth\beta >0$, and from
\begin{equation}
  \varepsilon_AR_A=\frak{s}{R_AD\over R_A-D}\,{1\over \coth\beta}\,,\end{equation}
we conclude that $\varepsilon_AR_A$ has the sign of $\frak{s}R_A(R_A-D)$, which is positive by definition of $\frak{s}$.

According to Eq.\ (\ref{eq119}) the sign of $\varepsilon_BR_B$ is opposite to the sign of $\frak{s}R_B(R_B+D)$, which is opposite to the sign of $\frak{s}R_A(R_A-D)$, according to Lemma 4, and consequently  $\varepsilon_BR_B$ has the sign of $\varepsilon_AR_A$ and is positive. \qed

\bigskip
According to Lemma \ref{lem5}, taking the square roots of $\varepsilon_AR_A$ and $\varepsilon_BR_B$, as done in Eq.\ (\ref{eq120}),  makes sense.

\begin{theorem}\label{th2} Let ${\cal A}$ (vertex $V_A$, radius $R_A$) be an emitter and ${\cal B}$ (vertex $V_B$, radius $R_B$) be a receiver at a distance $D=\overline{V_AV_B}$. 
Under the assumption $-1<J<0$,  and under the previous definitions, the field transfer from ${\cal A}$ to ${\cal B}$, which is expressed by Eq.\ (\ref{eq1}), can then be expressed, with reduced field amplitudes $u_A$ and $u_B$, in the form
\begin{equation}
  u_B(\vec \rho  \primespe)=\E^{-\frak{s}\beta}\,{\cal K}_{\frak{s}\beta}[u_A](\vec\rho  \primespe)\,,\hskip .5cm \mbox{if}\;\; D>0 \,,\end{equation}
or
\begin{equation}
  u_B(\vec \rho  \primespe)=\E^{-\frak{s}\beta}{\cal K}_{\frak{s}\beta}[\widetilde u_A](\vec\rho  \primespe)
  =\E^{-\frak{s}\beta}\,{\cal P}\circ{\cal K}_{\frak{s}\beta}[u_A](\vec\rho  \primespe)
  \,,\hskip .5cm \mbox{if}\;\; D<0 \,,\end{equation}
where ${\cal K}_{\frak{s}\beta}$ denotes the hyperbolic fractional Fourier transformation of the second kind and order $\frak{s}\beta$, and where $\widetilde u_A$ denotes the function $u_A$ symmetrized (\,$\widetilde u_A={\cal P}[u_A]$).
\end{theorem}

\noindent{\its Proof.}

\medskip
\noindent{\its i. Derivation of $\D\vec r/\lambda D$.}
We have
\begin{equation}
  {\D\vec r\over \lambda D}
  ={\varepsilon_AR_A\over D}\,\D\vec\rho\,.
\end{equation}
Then
\begin{eqnarray}
  \left({\varepsilon_AR_A\over D}\right)^2={{R_A}^{\! 2}\over (R_A-D)^2}\tanh^2\beta
  \rap &=&\rap -{{R_A}^{\! 2}\over R_A-D}\,{R_B+D\over D(D-R_A+R_B)}\nonumber \\
  \rap &=&\rap -{R_A\over R_B}\;{R_B+D\over R_A-D}\;{R_AR_B\over D(D-R_A+R_B)}\nonumber \\
  \rap &=&\rap {\varepsilon_AR_A\over \varepsilon_BR_B}\;{R_AR_B\over D(D-R_A+R_B)}\,.
\end{eqnarray}
From
\begin{equation}
  {\sinh^2\beta\over \cosh^2\beta}=\tanh^2\beta ={-(R_A-D)(R_B+D)\over D(D-R_A+R_B)}\,,\end{equation}
we deduce
\begin{equation}
  {1\over \cosh^2\beta}={(R_A-D)(R_B+D) +D(D-R_A+R_B)
    \over  D(D-R_A+R_B)}={R_AR_B\over  D(D-R_A+R_B)}\,,
\end{equation}
so that
\begin{equation}
  \left({\varepsilon_AR_A\over D}\right)^2= {1\over \cosh^2\beta}\;{\varepsilon_AR_A\over \varepsilon_BR_B}\,.
\end{equation}
Since $\varepsilon_AR_A>0$ and $\varepsilon_BR_B>0$, we obtain
\begin{equation}
  {\D\vec r\over \lambda D}={\varepsilon_AR_A\over D}\,\D\vec\rho={1\over \cosh\beta}\,\sqrt{{\varepsilon_AR_A\over \varepsilon_BR_B}}\,\D\vec\rho\,, \hskip .5cm \mbox{if}\;\;\; D>0\,,
\end{equation}
and
\begin{equation}
  {\D\vec r\over \lambda D}={\varepsilon_AR_A\over D}\,\D\vec\rho=-{1\over \cosh\beta}\,\sqrt{{\varepsilon_AR_A\over \varepsilon_BR_B}}\,\D\vec\rho\,, \hskip .5cm \mbox{if}\;\;\; D<0\,.
\end{equation}

\medskip
\noindent{\its ii. Derivation of $\vec r\vec\cdot\vec r'/\lambda D$.}
We have
\begin{equation}
  {\vec r\vec\cdot\vec r'\over\lambda D} = {1\over D}\sqrt{\varepsilon_AR_A\varepsilon_BR_B}\, \vec \rho\vec\cdot\vec \rho' = {1\over D}\sqrt{{-D^2R_AR_B\over (R_A-D)(R_B+D)}\tanh^2\beta} \, \vec \rho\vec\cdot\vec \rho'\,.\end{equation}
We use Eq.\ (\ref{eq116}) and, since $\coth^2\beta =-1/J$, we obtain
\begin{eqnarray}
  {\vec r\vec\cdot\vec r'\over\lambda D} ={1\over D}\sqrt{D^2\tanh^2\beta \left({1\over \tanh^2\beta}-1\right)}\, \vec \rho\vec\cdot\vec \rho'
  \rap &=&\rap {1\over D}\sqrt{D^2(1-\tanh^2\beta)}\,\vec \rho\vec\cdot\vec \rho'  \nonumber \\
  \rap &=&\rap {1\over D}\sqrt{{D^2\over \cosh^2\beta}}\,\vec \rho\vec\cdot\vec \rho'\,,
\end{eqnarray}
that is,
\begin{equation}
  {\vec r\vec\cdot\vec r'\over\lambda D}={\vec \rho\vec\cdot\vec \rho \primespe \over\cosh\beta} \,,\hskip .5cm \mbox{if}\;\;\; D>0\,,\end{equation}
and
\begin{equation}
  {\vec r\vec\cdot\vec r'\over\lambda D}=-{\vec \rho\vec\cdot\vec \rho \primespe \over\cosh\beta} \,,\hskip .5cm \mbox{if}\;\;\; D<0\,.\end{equation}

\medskip
\noindent{\its iii. Derivation of quadratic-phase factors.}
We begin with
\begin{equation}
  {1\over \lambda}\left({1\over D}-{1\over R_A}\right)r^2={1\over \lambda}{R_A-D\over R_AD}\lambda\varepsilon_AR_A\rho^2
  =\frak{s}\rho^2\tanh\beta\,.
\end{equation}
For the other factor, we have
\begin{equation}
  {1\over \lambda}\left({1\over R_B}+{1\over D}\right)r'^2={1\over \lambda}{R_B+D\over R_BD}\lambda\varepsilon_BR_B\rho'^2
  =-\frak{s}\rho^2\tanh\beta\,.
\end{equation}

\medskip
\noindent{\its iv. Integral.}
If $D>0$, Eq.\ (\ref{eq1}) is written
\begin{eqnarray}
  U_B\left(\sqrt{\lambda \varepsilon_BR_B}\vec \rho  \primespe\right)\rap &=&\rap {\I\over \cosh\beta}\,\sqrt{{\varepsilon_AR_A\over \varepsilon_BR_B}}
  \E^{\I{\frak s}\pi\rho'^2\tanh\beta} \\
  & &\hskip .3cm \times \;\;\int_{{\mathbb R}^2}  \E^{-\I{\frak s}\pi\rho^2\tanh\beta}
  \,\exp\left({2\I\pi\over\cosh\beta}\,\vec\rho\vec\cdot\vec\rho  \primespe\right)\,U_A \left(\sqrt{\lambda \varepsilon_BR_B}\vec \rho \right)\,\D\vec\rho\,, \nonumber
  \end{eqnarray}
that is
\begin{eqnarray}
  u_B(\vec \rho  \primespe)\rap &=&\rap{\I\over \cosh\beta}
   \,\E^{\I{\frak s}\pi\rho'^2\tanh\beta}
   \,\int_{{\mathbb R}^2}  \E^{-\I{\frak s}\pi\rho^2\tanh\beta}\exp\left({2\I\pi\over\cosh\beta}\,\vec\rho\vec\cdot\vec\rho '\right)\,u_A(\vec \rho )\,\D\vec\rho\nonumber \\
   \rap &=&\rap \E^{-\frak{s}\beta}\,{\cal K}_{\frak{s}\beta}[u_A](\vec\rho  \primespe)\,.
\end{eqnarray}

If $D<0$, we obtain
\begin{eqnarray}
  U_B\left(\sqrt{\lambda \varepsilon_BR_B}\vec \rho '\right)\rap &=&\rap {-\I\over \cosh\beta}\,\sqrt{{\varepsilon_AR_A\over \varepsilon_BR_B}}
  \E^{\I{\frak s}\pi\rho'^2\tanh\beta} \\
  & &\hskip .3cm \times \;\;\int_{{\mathbb R}^2}  \E^{-\I{\frak s}\pi\rho^2\tanh\beta}
  \,\exp\left(-{2\I\pi\over\cosh\beta}\,\vec\rho\vec\cdot\vec\rho '\right)\,U_A \left(\sqrt{\lambda \varepsilon_BR_B}\vec \rho \right)\,\D\vec\rho\,, \nonumber
  \end{eqnarray}
that is
\begin{eqnarray}
  u_B(\vec \rho  \primespe)\rap &=&\rap {-\I\over \cosh\beta}
   \E^{\I{\frak s}\pi\rho'^2\tanh\beta}
   \,\int_{{\mathbb R}^2}  \E^{-\I{\frak s}\pi\rho^2\tanh\beta}\exp\left(-{2\I\pi\over\cosh\beta}\,\vec\rho\vec\cdot\vec\rho  \primespe\right)\,u_A(\vec \rho )\,\D\vec\rho\nonumber \\
   &=&\rap {\I\over \cosh\beta}
   \E^{\I{\frak s}\pi\rho'^2\tanh\beta}
   \,\int_{{\mathbb R}^2}  \E^{-\I{\frak s}\pi\rho^2\tanh\beta}\exp\left({2\I\pi\over\cosh\beta}\,\vec\rho\vec\cdot\vec\rho '\right)\,u_A(-\vec \rho )\,\D\vec\rho\nonumber \\
   \rap &=&\rap \E^{-\frak{s}\beta}\,{\cal K}_{\frak{s}\beta}[\widetilde u_A](\vec\rho  \primespe)
   =  \E^{-\frak{s}\beta}\,{\cal K}_{\frak{s}\beta}\circ{\cal P}[u_A](\vec\rho  \primespe)
   =  \E^{-\frak{s}\beta}\,{\cal P}\circ{\cal K}_{\frak{s}\beta}[u_A](\vec\rho  \primespe)\,.
\end{eqnarray}
The proof is complete. \qed

\section{Application to the refracting spherical cap}\label{sect5}

The advantage of expressing diffraction phenomena by means of fractional-order Fourier trans\-for\-ma\-tions---whether circular or hyperbolic---is that certain problems  can be addressed simply by  manipulating the fractional orders, without the need to explicitly write out complete integral expressions. This is because a fractional transformation is fully determined by its order, once its kind is known, and writing the corresponding integral does not provide additional information.   This method has been applied to diffraction problems involving circular fractional-order Fourier transformations \cite{PPF3,PPF1} and we propose extending it to diffraction phenomena associated with hyperbolic fractional transformations of the first and the second kind.

One particular question involves imaging through refracting caps \cite{PPF3,PPF1}. Whereas the  properties of refracting caps may be derived from the fundamental laws of geometrical optics (Fermat's principle, Snell's law), deriving them  from diffraction theory affords the integration of  paraxial geometrical optics
into the electromagnetic wave theory, and has an interest for elaborating a unitary theory of optics.

We will apply previous results, particularly the composition law of hyperbolic fractional-order Fourier transformations, to coherent geometrical imaging by a refracting spherical cap  \cite{PPF2}. By coherent imaging, we mean that the field amplitude of the image ${\cal A}'$ is equal to the field amplitude of the object ${\cal A}$, including the phase, up to a magnification factor (denoted as $m$) and a multiplicative constant factor, that is,
\begin{equation}
  U_{A'}(\vec r')= {1\over m}\,U_A\left({\vec r'\over m}\right)\,.\label{eq146}\end{equation}
(In Eq. (\ref{eq146}), $m$ is the lateral magnification for the object and image positions with respect to the refracting surface. The factor $1/m$, before $U_A$, is necessary for power conservation.) The imaging is geometrical, because we do not take into account the diffraction effects due to a limited  aperture of the refracting cap.

The basic results of coherent imaging by a refracting spherical cap are \cite{PPF1,PPF3,GB1,GB2,PPF2}:
\begin{enumerate}
\item[$\bullet$] {\its Double conjugation.} If ${\cal A}'$ is the coherent geometrical image of ${\cal A}$, formed by a refracting spherical cap, then the vertex of ${\cal A}'$ is the conjugate point---in the sense used in  paraxial optics---of the vertex of ${\cal A}$, and the center of curvature of ${\cal A}'$ is the conjugate point of the center of ${\cal A}$.
 (The conjugation of centers of curvature is characteristic of coherent imaging, as it results from the preservation of phases in the imaging process.)
\item[$\bullet$] {\its Conjugation formula  and corresponding lateral magnification (in accordance with paraxial optics).}
  Explicitely, if $R_D= \overline{VC}$ is the radius  of curvature of the refracting cap (vertex $V$, center $C$), if $n$ and $n'$ are the refractive indices  of the object and image spaces, and if $d$ is the algebraic measure from $V$  to the object vertex $V_A$ ($d=\overline{VV_A}$), and $d'$ from $V$ to the image vertex $V_{A'}$ ($d'=\overline{VV_{A'}}$), the conjugation formula for vertices is
  \begin{equation}{n'\over d}={n\over d}+{n'-n\over R_D}\,,\label{eq140}\end{equation}
 and  the corresponding lateral magnification (at vertices) is
  \begin{equation}
    m_{\rm v}={nd'\over n'd}\,.
    \end{equation}

\item[$\bullet$] {\its Radius-magnification law (Bonnet's law).} If $m_{\rm v}$ is the lateral magnification at vertices and $m_{\rm c}$ the lateral magnification at centers of curvature (between the object and its image), the radius-magnification law is
  \begin{equation}
    m_{\rm r}={R_{A'}\over R_A}={n'\over n}m_{\rm v}m_{\rm c}\,,\end{equation}
  where $R_A$ is the radius of curvature of the (spherical) object ${\cal A}$, and $R_{A'}$ the radius  of curvature of the image ${\cal A}'$.
(The concept of radius magnification may be seen as generalizing  the differential longitudinal magnification, used in paraxial optics, to finite segments.)
\end{enumerate}

Those findings can be independently obtained within the framework of geometrical optics \cite{PPF2}, or in the metaxial-optics theory\cite{GB1,GB2,PPF1}, or by composing two circular fractional-order Fourier transformations \cite{PPF3,PPF1}. Here, we will prove these results by composing two hyperbolic fractional-order Fourier transformations of the same kind.

A preliminary remark is necessary. Let us consider a spherical refracting cap ${\cal D}$ separating two homogeneous and isotropic propagation media, and splitting the physical space into the object and image spaces. Let ${\cal A}$ be an emitter in the object space and ${\cal A}'$ be a receiver in the image space. The field transfer from ${\cal A}$ to ${\cal A}'$ is seen as the composition of two fractional-order Fourier transformations: one from ${\cal A}$ to ${\cal D}$, the other from ${\cal D}$ to ${\cal A}'$. Each transformation may be a circular or a hyperbolic transformation. We have a priori to examine the following products ($\alpha$ and $\beta$ refer to transfers from ${\cal A}$ to ${\cal D}$, and $\alpha '$ and $\beta '$ from ${\cal D}$ to ${\cal A}'$):
\begin{enumerate}
\item ${\cal F}_{\alpha '}\circ{\cal F}_\alpha$ ;  ${\cal H}_{\beta '}\circ{\cal H}_\beta$ ; ${\cal K}_{\beta '}\circ{\cal K}_\beta$ ;
\item ${\cal H}_{\beta '}\circ{\cal F}_\alpha$ ; ${\cal K}_{\beta '}\circ{\cal F}_\alpha$ ; ${\cal F}_{\alpha '}\circ {\cal H}_{\beta }$ ;  ${\cal F}_{\alpha '}\circ {\cal K}_{\beta }$ ;
\item ${\cal K}_{\beta '}\circ{\cal H}_\beta$ ; ${\cal H}_{\beta '}\circ{\cal K}_\beta$ .
\end{enumerate}

The field transfer from ${\cal A}$ to ${\cal A}'$ is an imaging if the composed transfer-operator is the identity operator ${\cal I}$, or the parity operator ${\cal P}$, up to a multiplicative factor. (For the parity operator, the imaging transforms $u_A$ into $u_{A'}=\widetilde u_A$, which may correspond to an image inverted with respect to the object, a frequent situation in optics.) Since
${\cal F}_{\alpha '}\circ{\cal F}_\alpha ={\cal F}_{\alpha '+\alpha}$, we obtain ${\cal F}_{\alpha '+\alpha}={\cal I}={\cal F}_0$, when $\alpha '+\alpha =0$, and ${\cal F}_{\alpha '+\alpha}={\cal P}={\cal F}_{\pm \pi}$, when $\alpha '+\alpha =\pm \pi$. Also, since ${\cal H}_0={\cal I}$, we can obtain ${\cal H}_{\beta '}\circ{\cal H}_\beta={\cal H}_{\beta '+\beta }={\cal I}$, when $\beta +\beta '=0$,  and  ${\cal K}_{\beta '}\circ{\cal K}_\beta=-\E^{2\beta '}\,{\cal H}_{\beta -\beta '}\circ{\cal P}=-\E^{2\beta '}\,{\cal P}$, when $\beta -\beta '=0$. But since  the operator ${\cal K}_{\beta '}\circ{\cal H}_\beta={\cal K}_{\beta '+\beta}$ is  proportional to neither ${\cal I}$ (even when $\beta '+\beta =0$) nor ${\cal P}$, and since  the operator ${\cal H}_{\beta '}\circ{\cal K}_\beta=\E^{2\beta '}\,{\cal K}_{\beta '-\beta}$ is also proportional to neither ${\cal I}$ nor  ${\cal P}$ (even when $\beta '=\beta$),  we cannot obtain the identity or the parity operator by composing a hyperbolic transformation of the first kind with one of the second kind.

In an article to be published \cite{PPF7}, we  prove that the composition of a circular fractional-order Fourier transformation with a hyperbolic transformation (as mentioned in item 2 above) can be neither the identity operator nor the parity operator, except in certain ``trivial'' cases of limited interest in optics. Consequently, only compositions of the type described in item 1 above can yield  the identity or the parity operator. Since the composition ${\cal F}_{\alpha '}\circ{\cal F}_\alpha$ has already been addressed in previous publications \cite{PPF3,PPF1}, in what follows we will  consider only the products ${\cal H}_{\beta '}\circ{\cal H}_\beta$ and  ${\cal K}_{\beta '}\circ{\cal K}_\beta$.

In the next sections, the field transfer from ${\cal A}$ to ${\cal D}$ will be  related to the parameter $J$, and the field transfer from ${\cal D}$ to ${\cal A}'$ to the parameter $J'$.

\subsection{The case $J<-1$ and $J'<-1$: product ${\cal H}_{\beta '}\circ{\cal H}_\beta$}\label{sect51}

Let ${\cal D}$ (vertex $V_D$,  center  of curvature $C_D$, radius $R_D=\overline{V_DC_D}$) be a spherical cap separating two propagation media of respective refractive indices $n$ and $n'$ (corresponding wavelengths are such that $n\lambda =n'\lambda '$). Let ${\cal A}$ (radius $R_A=\overline{V_AC_A}$) be a spherical emitter in the object space (index $n$) and ${\cal A}'$  (radius $R_{A'}=\overline{V_{A'}C_{A'}}$) a spherical cap in the image space (see Fig.\ \ref{fig2}). For the distances between ${\cal D}$ and ${\cal A}$ (emitter  or object)  and ${\cal A}'$ (receiver or image), we refer to the usual definitions of geometrical optics:  $d$ is the algebraic measure from ${\cal D}$ to ${\cal A}$, that is, $d=\overline{V_DV_A}$, and $d\primespe$  is that from ${\cal D}$ to ${\cal A}'$, i.e. $d\primespe =\overline{V_DV_{A'}}$.

\begin{figure}
  \centering
    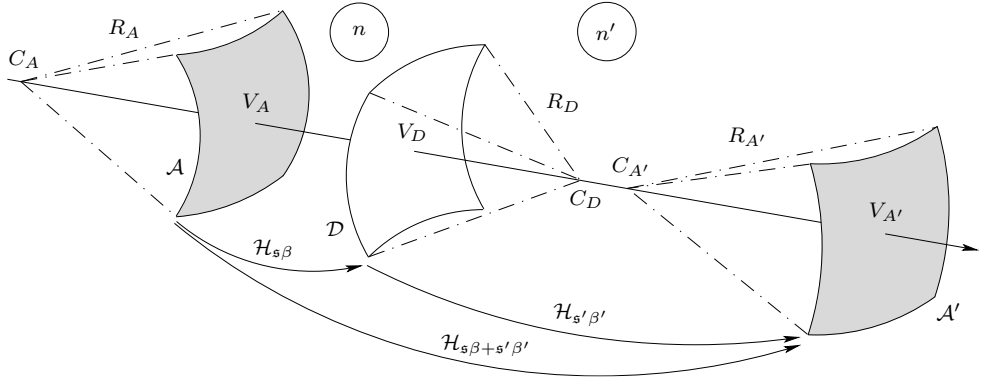
    \caption{\small Elements for representing the field transfer from a spherical emitter ${\cal A}$ to a spherical receiver ${\cal A}'$ through the refracting cap ${\cal D}$. Light propagates from left to right.\label{fig2}}
\end{figure}

According to Theorem \ref{th1}, we assume the field transfer from ${\cal A}$ to ${\cal D}$ to  be represented by a hyperbolic fractional Fourier transform of the first kind with order $\frak{s}\beta$, namely  ${\cal H}_{\frak{s}\beta}$, and the field transfer from ${\cal D}$ to ${\cal A}'$ by a transformation of the first kind of order $\frak{s}'\beta '$, namely ${\cal H}_{\frak{s}'\beta '}$ (see Remark \ref{rem7}). Since the reduced field amplitudes on ${\cal D}$ are the same for both transfers (as we shall show, see Sect.\ \ref{sect512}), the field transfer from ${\cal A}$ to ${\cal A}'$ will then be described as the composition of the two previous transformations (see Fig.\ \ref{fig2}), which results in
\begin{equation}
  u_{A'}=\frak{s'}\frak{s}\,\E^{\frak{s'}\beta '+\frak{s}\beta}\,{\cal H}_{\frak{s'}\beta '+\frak{s}\beta}[u_A]\,,\label{eq142}\end{equation}
or
\begin{equation}
  u_{A'}=\frak{s'}\frak{s}\,\E^{\frak{s'}\beta '+\frak{s}\beta}\,{\cal P}\circ{\cal H}_{\frak{s'}\beta '+\frak{s}\beta}[u_A]\,.\label{eq143}\end{equation}
The composition will make sense, indeed, only if reduced variables on ${\cal D}$ are the same for both transfers. Then ${\cal A}'$ will be the image of ${\cal A}$ if $\frak{s}\beta +\frak{s}'\beta ' =0$. We  now examine the conditions for that.

\subsubsection{Transfers to be composed}
We adapt the notation of Sect.\ \ref{sect4}: for describing the field transfer from ${\cal A}$ to ${\cal D}$, we replace the receiver ${\cal B}$ with ${\cal D}$, the parameter $\varepsilon_B$ with $\varepsilon_D$, the reduced amplitude $u_B$ with $u_D$, etc.
In Section \ref{sect4}, the distance $D$ is taken from the emitter ${\cal A}$ to the receiver ${\cal B}$. Here the distance $d$ is taken from the receiver ${\cal D}$ to the emitter ${\cal A}$, so that the results of Sect.\ \ref{sect4} can be applied for expressing the transfer from ${\cal A}$ to ${\cal D}$ if  $D=-d$. The order $\beta$ is thus defined by
  \begin{equation}
    \coth^2\beta =-J=-{(R_A+d)(R_D-d)\over d(d+R_A-R_D)}>1\,,\hskip 1cm  -\beta d>0\,.\end{equation}
  We use
  \begin{equation}
    \varepsilon_A=-\frak{s}{d\over R_A+d}\,\coth\beta\,,\hskip 1cm  \varepsilon_D=-\frak{s}{d\over d-R_D}\,\coth\beta\,,
  \end{equation}
  where $\frak{s}$ denotes the sign of $R_A(R_A+d)$, which is also the sign of $R_D(R_D-d)$.
  
 If $\vec r$ and $\vec s$ are spatial variables on ${\cal A}$ and ${\cal D}$, reduced variables are
  \begin{equation}
    \vec \rho={\vec r\over \sqrt{\lambda \varepsilon_AR_A}}\,,\hskip 1cm  \vec \sigma={\vec s\over \sqrt{\lambda \varepsilon_DR_D}}\,.
  \end{equation}
   According to Theorem \ref{th1} (and Remark \ref{rem7}), the field transfer from ${\cal A}$ to ${\cal D}$ takes the form
  \begin{equation}
    u_{D_-}(\vec \sigma )=\frak{s}\,\E^{-\frak{s}\beta}\,{\cal H}_{\frak{s}\beta}[u_A](\frak{s}\vec\sigma )\,,\label{eq148}
  \end{equation}
  where $u_{D_-}$ denotes the reduced amplitude of the field incident on ${\cal D}$ (just before refraction).

  For the field transfer from ${\cal D}$ to ${\cal A}'$, we replace ${\cal A}$ in Sect.\ \ref{sect4} with ${\cal D}$ (the emitter) and ${\cal B}$ with ${\cal A}'$ (the receiver). We also replace $\varepsilon_A$ with $\varepsilon_D'$, $\varepsilon_{B}$ with $\varepsilon_{A'}$ and $D'$ with $d\primespe$. The order $\beta '$ is such that
   \begin{equation}
    \coth^2\beta '=-J'={(R_D-d\primespe)(R_{A'}+d\primespe)\over d\primespe(d\primespe-R_D+R_D)}>1\,,\hskip 1cm  \beta ' d\primespe>0\,,\end{equation}
and
  \begin{equation}
    \varepsilon_D'=\frak{s}'{d\primespe\over R_D-d\primespe}\,\coth\beta '\,,\hskip 1cm  \varepsilon_{A'}=\frak{s}'{d\primespe\over R_{A'}+d\primespe}\,\coth\beta '\,,
  \end{equation}
  where $\frak{s}'$ denotes the sign of $R_D(R_D-d\primespe)$, which is also the sign of $R_{A'}(R_{A'}+d\primespe)$.

  Spatial variables are $\vec s$ and $\vec r'$ on ${\cal D}$ and ${\cal A}'$, and reduced variables are 
  \begin{equation}
    \vec \sigma '={\vec s\over \sqrt{\lambda '\varepsilon_D'R_D}}\,,\hskip 1cm  \vec \rho '={\vec r'\over \sqrt{\lambda '\varepsilon_{A'}R_{A'}}}\,.
  \end{equation}
The field transfer from ${\cal D}$ to ${\cal A}'$ takes the form
  \begin{equation}
    u_{A'}(\vec \rho ' )=\frak{s}'\,\E^{-\frak{s}'\beta '}\,{\cal H}_{\frak{s}'\beta'}[u_{D_+}](\frak{s}'\vec\rho ' )\,,
   \label{eq152} \end{equation}
  where $u_{D_+}$ denotes the reduced amplitude of the field emerging from ${\cal D}$ (just after refraction).

\subsubsection{Reduced field amplitudes on ${\cal D}$}\label{sect512}

Boundary conditions at the refracting surface imply that the tangential component of the electric field  is continuous at the interface. Since we develop a scalar theory, we conclude that for every point $\vec s$ of ${\cal D}$ we have
$U_{D_-}(\vec s)=U_{D_+}(\vec s)$, where $U_{D_-}$ is the field amplitude on ${\cal D}$ before refraction, and $U_{D_+}$ after refraction. We may denote $U_D\equiv U_{D_-}\equiv U_{D_+}$.

The corresponding reduced amplitudes are
\begin{equation}
  u_{D_-}(\vec\sigma )=\sqrt{\lambda\varepsilon_DR_D}\;U_D\left(\sqrt{\lambda \varepsilon_DR_D}\,\vec\sigma\right)\,,\label{eq150a}\end{equation}
and
\begin{equation}
  u_{D_+}(\vec\sigma ')=\sqrt{\lambda '\varepsilon_D'R_D}\;U_D\left(\sqrt{\lambda '\varepsilon_D'R_D}\,\vec\sigma '\right)\,,\label{eq151a}\end{equation}
from which we deduce
\begin{equation}
  {1\over \sqrt{\lambda\varepsilon_DR_D}}\; u_{D_-}\!\!\left({\vec s\over \sqrt{\lambda\varepsilon_DR_D}}\right) =U_D(\vec s )={1\over \sqrt{\lambda'\varepsilon_D'R_D}}\; u_{D_+}\!\!\left({\vec s\over \sqrt{\lambda '\varepsilon_D'R_D}}\right)\,.
  \end{equation}

To describe the field transfer from ${\cal A}$ to ${\cal A}'$, we compose $u_{A}\longmapsto u_{D_-}$ with $u_{D_+}\longmapsto u_{A'}$, which makes sense only if reduced variables on ${\cal D}$ for both mappings are identical to each other, which gives 
  \begin{equation}
    \lambda '\varepsilon_D'R_D = \lambda \varepsilon_DR_D\,,\label{eq156}
  \end{equation}
  so that $\vec \sigma '=\vec \sigma$ and  $u_{D_-}\equiv u_{D_+}$.

  \subsubsection{Compositions for imaging}\label{sect513}
  We compose the mappings given by Eqs.\ (\ref{eq148}) and (\ref{eq152}). More precisely, since $u_{D_-}\equiv u_{D_+}$, we conclude:
  \begin{itemize}
  \item[$\bullet$] If $\frak{s'}=\frak{s}=1$, we have $u_{D_-}(\vec \sigma )=\E^{-\beta}\,{\cal H}_\beta[u_A](\vec\sigma)$ and $u_{A'}(\vec \rho')=\E^{-\beta '}\,{\cal H}_{\beta '}[u_{D_+}](\vec\rho ')$, so that
    $u_{A'}(\vec \rho')=\E^{-\beta -\beta '}\,{\cal H}_{\beta +\beta '}[u_{A}](\vec\rho ')$. The spherical cap ${\cal A}'$ is the coherent image of ${\cal A}$ if $\beta '=-\beta$. Then $u_{A'}\equiv u_A$, and reduced variables on ${\cal A}$ and ${\cal A}'$ are identical: $\vec\rho '=\vec\rho$.
 \goodbreak 
     \item[$\bullet$] If $\frak{s'}=\frak{s}=-1$, we have $u_{D_-}(\vec \sigma )=-\E^{\beta}\,{\cal P}\circ{\cal H}_{-\beta}[u_A](\vec\sigma)$ and $u_{A'}(\vec \rho')=-\E^{\beta '}\,{\cal P}\circ{\cal H}_{-\beta '}[u_{D_+}](\vec\rho ')$, so that
       $u_{A'}(\vec \rho')=\E^{\beta+\beta '}\,{\cal H}_{-\beta -\beta '}[u_{A}](\vec\rho ')$. The spherical cap ${\cal A}'$ is the coherent image of ${\cal A}$ if $\beta '=-\beta$. Then $u_{A'}\equiv u_A$, and reduced variables on ${\cal A}$ and ${\cal A}'$ are identical: $\vec\rho '=\vec\rho$.
     \item[$\bullet$] If $\frak{s'}=-\frak{s}=-1$, we have $u_{D_-}(\vec \sigma )=\E^{-\beta}\,{\cal H}_{\beta}[u_A](\vec\sigma)$ and $u_{A'}(\vec \rho')=-\E^{\beta '}\,{\cal P}\circ{\cal H}_{-\beta '}[u_{D_+}](\vec\rho ')$, so that
          $u_{A'}(\vec \rho')=-\E^{\beta '-\beta }\,{\cal P}\circ{\cal H}_{\beta -\beta '}[u_{A}](\vec\rho ')$. The spherical cap ${\cal A}'$ is the coherent image of ${\cal A}$ if $\beta '=\beta$. Then $u_{A'}\equiv -{\cal P}[u_A]=-\widetilde u_A$, and reduced variables on ${\cal A}$ and ${\cal A}'$ are opposite: $\vec\rho '=-\vec\rho$.
           \item[$\bullet$] If $\frak{s'}=-\frak{s}=1$, we have $u_{D_-}(\vec \sigma )=-\E^{\beta}\,{\cal P}\circ{\cal H}_{-\beta}[u_A](\vec\sigma)$ and $u_{A'}(\vec \rho')=\E^{-\beta '}\,{\cal H}_{\beta '}[u_{D_+}](\vec\rho ')$, so that
     $u_{A'}(\vec \rho')=-\E^{\beta -\beta' }\,{\cal P}\circ{\cal H}_{\beta' -\beta }[u_{A}](\vec\rho ')$. The spherical cap ${\cal A}'$ is the coherent image of ${\cal A}$ if $\beta '=\beta$. Then $u_{A'}\equiv -{\cal P}[u_A]=-\widetilde u_A$, and reduced variables on ${\cal A}$ and ${\cal A}'$ are opposite: $\vec\rho '=-\vec\rho$.
\end{itemize}

\subsubsection{Conjugation formula}
 
Equation (\ref{eq156}) leads to
  \begin{equation}
    {n'\over \varepsilon_D'}={n\over \varepsilon_D}\,,\label{eq56}\end{equation}
  and then
   \begin{equation}
     \frak{s}'n'{R_D-d'\over d\primespe\coth\beta '}=-\frak{s} n{R_D-d\over d\coth\beta}\,.\label{eq155}\end{equation}

   The spherical cap  ${\cal A}'$ is the image of ${\cal A}$, if  $\frak{s}\beta +\frak{s}'\beta '=0$. Then  $\frak{s}'/\coth\beta '=-\frak{s}/\coth\beta$, and Eq.\ (\ref{eq155}) gives
     \begin{equation}
       {n'\over d\primespe}-{n'\over R_D}={n\over d}-{n\over R_D}\,,\label{eq59}\end{equation}
     that is
     \begin{equation}
       {n'\over d\primespe}={n\over d}+{n'-n\over R_D}\,,\label{eq60}\end{equation}
     which is a conjugation formula for the refracting sphere. It indicates that vertices of ${\cal A}$ and ${\cal A}'$ are conjugate points.

 \subsubsection{Lateral magnification}\label{sect515}}

 The lateral magnification between conjugate points $V_{A'}$ and $V_A$ is $m_{\rm v}$, with $\vec r'=m_{\rm v}\vec r$.
 According to Sect.\ \ref{sect513},  reduced variables on $u_A$ and $u_{A'}$ are identical, up to sign: $\vec \rho '=\pm \vec \rho$. Then
 \begin{equation}
   \vec r'=\sqrt{\lambda '\varepsilon_{A'}R_{A'}}\,\vec \rho ' =\pm \sqrt{\lambda '\varepsilon_{A'}R_{A'}}\,\vec \rho = \pm {\sqrt{\lambda '\varepsilon_{A'}R_{A'}}\over \sqrt{\lambda \varepsilon_{A}R_{A}}}\,\vec r\,,\end{equation}
 and
   \begin{equation}
     {m_{\rm v}}=\pm{\sqrt{\lambda '\varepsilon_{A'}R_{A'}}\over \sqrt{\lambda \varepsilon_{A}R_{A}}}\,.\label{eq161}\end{equation}
From $\frak{s} '\beta '=-\frak{s}\beta$, we deduce  $\frak{s} '\coth\beta '=-\frak{s}\coth\beta$, and we obtain
     \begin{equation}
       {m_{\rm v}}^{\! 2}={\lambda '\varepsilon_{A'}R_{A'}\over \lambda \varepsilon_{A}R_{A}}
       =-{\frak{s}'\over\frak{s}}\,{nd'\over n'd}\,{R_{A'}(R_A+d)\over R_A (R_{A'}+d')}\,{\coth\beta '\over \coth\beta}={nd'\over n'd}\,{R_{A'}(R_A+d)\over R_A (R_{A'}+d')}\,.\label{eq64}
     \end{equation}
 From $\coth^2\beta =-J$ and from Eq.\ (\ref{eq26}),  we deduce
 \begin{equation}
   {R_AR_D\over (R_A+d)(R_D-d)}=1+{1\over J}=1-{1\over \coth^2\beta}={1\over \cosh^2\beta}\,.
 \end{equation}
 Similarly we write, for the transfer from ${\cal D}$ to ${\cal A}'$
  \begin{equation}
   {R_DR_{A'}\over (R_D-d')(R_{A'}+d')}=1+{1\over J'}=1-{1\over \coth^2\beta '}={1\over \cosh^2\beta'}\,.
  \end{equation}
  Since $\beta '=\pm\beta$, we obtain
   \begin{equation}
     {R_AR_D\over (R_A+d)(R_D-d)}= {R_DR_{A'}\over (R_D-d')(R_{A'}+d')}\,,\label{eq67}
   \end{equation}
   and Eq.\ (\ref{eq64}) provides
   \begin{equation}
     {m_{\rm v}}^{\!2}={nd'\over n'd}\,{R_D-d\primespe\over R_D-d}\,.
   \end{equation}
   From Eq.\ (\ref{eq59}), we deduce
   \begin{equation}
     n'\,{R_D-d\primespe\over d\primespe}=n\,{R_D-d\over d}\,,\label{eq166}\end{equation}
   so that
    \begin{equation}
      {m_{\rm v}}^{\!2}={n^2 d\primespe^2\over n'^2d^2}\label{eq169}\,.
    \end{equation}
Eventually, we have $-\beta d>0$ and $\beta 'd\primespe>0$, so that
    \begin{enumerate}
    \item[$\bullet$] if $\frak{s'}=\frak{s}$,  then $\beta '=-\beta$ and  $d\primespe/d>0$.
    \item[$\bullet$]  if $\frak{s'}=-\frak{s}$,  then $\beta '=\beta$ and $d\primespe/d<0$.
    \end{enumerate}
    But according to Sect.\ \ref{sect513}, if  $\frak{s'}=\frak{s}$, then  $\vec \rho \primespe =\vec \rho$, and $ m_{\rm v}>0$ ;  if  $\frak{s'}=-\frak{s}$, then  $\vec \rho \primespe =-\vec \rho$, and $ m_{\rm v}<0$. We conclude that $m_{\rm v}$ has the sign of $d'/d$, so that Eq.\ (\ref{eq169}) leads to 
     \begin{equation}
       m_{\rm v}={nd\primespe\over n'd}\,,\label{eq168}\end{equation}
     which is the expression of the lateral magnification for the refracting sphere in paraxial optics. 

   \subsubsection{Conjugation of centers of curvature}\label{sect413}

   If $C_A$ denotes the center of curvature of ${\cal A}$ and $C_{A'}$ that of ${\cal A}'$, we denote $q=\overline{V_DC_A}=d+R_A$ and $q\primespe =\overline{V_DC_{A'}}=d\primespe +R_{A'}$. From Eq.\ (\ref{eq67})  we deduce
   \begin{equation}
     {q-d\over q}\,{1\over R_D-d}={q\primespe -d\primespe\over q\primespe}\,{1\over R_D-d\primespe}\,,
   \end{equation}
   which gives, according to Eq.\ (\ref{eq166}),
   \begin{equation}
     n{q-d\over qd}=n'{q\primespe -d\primespe\over d\primespe q\primespe}\,,\label{eq170}
   \end{equation}
   that is, according to Eq.\ (\ref{eq60}),
   \begin{equation}
     {n'\over q\primespe}-{n\over q}={n'\over d\primespe}-{n\over d}={n'-n\over R_D}\,,\end{equation}
   so that
   \begin{equation}
     {n'\over q\primespe}={n\over q}+{n'-n\over R_D}\,.\label{eq171}\end{equation}
   Equation (\ref{eq171}) shows that $C_A$ and $C_{A'}$ are conjugate points.

   According to Eq.\ (\ref{eq168}) the lateral magnification at centers of curvature is $m_{\rm c}$ with
     \begin{equation}
       m_{\rm c}={nq\primespe \over n'q}\,.\label{eq172}\end{equation}
    
     \subsubsection{Radius-magnification law}\label{sect414}
     The radius magnification between the spherical cap ${\cal A}$ and its coherent image ${\cal A}'$ is defined by
     \begin{equation}
       m_{\rm r}={R_{A'}\over R_A}\,.
     \end{equation}
      We use Eq.\ (\ref{eq170}) and derive
     \begin{equation}
       m_{\rm r}={q'-d'\over q-d}={nq'd'\over n'qd}={n'\over n}\,m_{\rm v}\,m_{\rm c}\,,
     \end{equation}
     which is the Bonnet's law of radius magnification \cite{PPF1}.

 \subsubsection{Field transfer from an object to its image}\label{sect518}
 We assume $\frak{s}\beta +\frak{s}'\beta '=0$, so that ${\cal A}'$ is the coherent image of ${\cal A}$.
 The composition of ${\cal H}_{\frak{s}\beta}$ and ${\cal H}_{\frak{s}'\beta '}$ is the identity or the parity operator (up to sign). We conclude that the reduced field amplitudes on an object ${\cal A}$ and its coherent image ${\cal A}'$  are such that
 \begin{equation}
   u_{A'}(\vec \rho \primespe)=u_{A}(\vec \rho\primespe )\,,\hskip .5cm \mbox{if}\;\;\; m_{\rm v}>0\,,\label{eq178}\end{equation}
  \begin{equation}
  u_{A'}(\vec \rho \primespe)=-\widetilde u_{A}(\vec \rho\primespe )= -u_{A}(-\vec \rho\primespe )\,,\hskip 0.5cm\mbox{if}\;\;\;  m_{\rm v}<0\,,\,.\label{eq179}\end{equation}
Then, when $m_{\rm v}>0$, \begin{eqnarray}
   U_{A'}(\vec r')\rap &=&\rap \sqrt{1\over\lambda '\varepsilon_{A'}R_{A'}}\,u_{A'}\left(\vec r'\over
   \sqrt{\lambda '\varepsilon_{A'}R_{A'}}\right)
   = \sqrt{1\over \lambda '\varepsilon_{A'}R_{A'}}\,u_{A}\left(\vec r'\over
   \sqrt{\lambda '\varepsilon_{A'}R_{A'}}\right)\nonumber \\ &=&\rap   \sqrt{\lambda \varepsilon_{A}R_{A}\over \lambda '\varepsilon_{A'}R_{A'}}\,U_A\bigg(\sqrt{\lambda \varepsilon_{A}R_{A}\over
     \lambda '\varepsilon_{A'}R_{A'}}\,\vec r'\bigg)\nonumber \\ &=&\rap
        {1\over m_{\rm v}}\,U_A\left({\vec r'\over m_{\rm v}}\right)\,.\label{eq184a}
 \end{eqnarray}
When $m_{\rm v}<0$, the following derivation leads to the same result
\begin{eqnarray}
   U_{A'}(\vec r')\rap &=&\rap \sqrt{1\over\lambda '\varepsilon_{A'}R_{A'}}\,u_{A'}\left(\vec r'\over
   \sqrt{\lambda '\varepsilon_{A'}R_{A'}}\right)
   = -\sqrt{1\over \lambda '\varepsilon_{A'}R_{A'}}\,u_{A}\left(-\vec r'\over
   \sqrt{\lambda '\varepsilon_{A'}R_{A'}}\right)\nonumber \\ &=&\rap   -\sqrt{\lambda \varepsilon_{A}R_{A}\over \lambda '\varepsilon_{A'}R_{A'}}\,U_A\bigg(-\sqrt{\lambda \varepsilon_{A}R_{A}\over
     \lambda '\varepsilon_{A'}R_{A'}}\,\vec r'\bigg)\nonumber \\ &=&\rap
        {1\over m_{\rm v}}\,U_A\left({\vec r'\over m_{\rm v}}\right)\,.\label{eq184b}
 \end{eqnarray}

 Equations (\ref{eq184a}) and (\ref{eq184b})  take the form of Eq.\ (\ref{eq146}) and represent the coherent-imaging relationship between the field amplitudes on ${\cal A}$ and ${\cal A}'$.

\subsection{The case $-1<J<0$ and $-1<J'<0$: product ${\cal K}_{\beta '}\circ{\cal K}_\beta$}\label{sect52}

The issue is that of Sect.\ \ref{sect5}, but under the assumption $-1<J<0$.

\subsubsection{Composition for imaging}
The results of Sect.\ \ref{sect512} are still valid. According to Theorem \ref{th2}, we have to compose 
\begin{equation}
  u_{A}\longmapsto u_{D_-}=\E^{-\frak{s}\beta}\,{\cal K}_{\frak{s}\beta}[u_A]\,,\;\;\; \mbox{if}\;\;d<0\;\; \mbox{(we recall that $d=-D$)}\,,\end{equation}
or
\begin{equation}u_{A}\longmapsto u_{D_-}=\E^{-\frak{s}\beta}\,{\cal P}\circ {\cal K}_{\frak{s}\beta}[u_A]\,,\;\;\; \mbox{if}\;\;d>0 \,,
\end{equation}
with
\begin{equation}
  u_{D_+}\longmapsto u_{A'}=\E^{-\frak{s'}\beta '}\,{\cal K}_{\frak{s'}\beta '}[u_{D_+}]\;\;\; \mbox{if}\;\;d\primespe >0\,,\end{equation}
or with
\begin{equation}u_{D_+}\longmapsto u_{A'}=\E^{-\frak{s'}\beta'}\,{\cal P}\circ {\cal K}_{\frak{s'}\beta'}[u_{D_+}]\;\;\; \mbox{if}\;\;d\primespe <0\,,
\end{equation}
According to Proposition \ref{prop4},
and since $u_{D_+}\equiv u_{D_-}$, the composition results in
\begin{equation}
  u_{A'}=-\E^{\frak{s'}\beta ' -\frak{s}\beta }\,{\cal H}_{\frak{s}\beta -\frak{s'}\beta '}[u_A]\,,\label{eq186}\end{equation}
or in
\begin{equation}
  u_{A'}=-\E^{\frak{s'}\beta ' -\frak{s}\beta }\,{\cal P}\circ {\cal H}_{\frak{s}\beta -\frak{s'}\beta '}[u_A]\,.\label{eq187}\end{equation}
The field amplitude on ${\cal A}'$ is the image of the field amplitude on ${\cal A}$ if
$\frak{s'}\beta ' -\frak{s}\beta=0$, that is 
\begin{itemize}
\item $\beta '=\beta$, if $\frak{s}'=\frak{s}$,
\item $\beta '=-\beta$, if $\frak{s}'=-\frak{s}$.
  \end{itemize}

\subsubsection{Conjugation formula}

Since the distance from ${\cal A}$ to ${\cal D}$ is $D=-d$, the field transfer from ${\cal A}$ to ${\cal D}$  is described with parameters
\begin{equation}
  \varepsilon_A=-\frak{s}{d\over R_A+d}\,\tanh\beta\,,\hskip 1cm   \varepsilon_D=\frak{s}{d\over R_D-d}\,\tanh\beta\,,\end{equation}
and reduced space variables on ${\cal A}$ and ${\cal D}$
\begin{equation}
  \vec \rho={\vec r\over\sqrt{\lambda\varepsilon_AR_A}}\,,\hskip 1cm
  \vec \sigma={\vec r\over\sqrt{\lambda\varepsilon_DR_D}}\,.
\end{equation}
The field transfer from ${\cal D}$ to ${\cal A}'$ is described with
\begin{equation}
  \varepsilon_D'=\frak{s'}{d\primespe \over R_A-d'}\,\tanh\beta\,,\hskip 1cm   \varepsilon_{A'}=-\frak{s'}{d\primespe \over R_{A'}+d\primespe}\,\tanh\beta '\,,\end{equation}
and reduced space variables on ${\cal D}$ and ${\cal A}'$
\begin{equation}
  \vec \sigma '={\vec r\over\sqrt{\lambda '\varepsilon_D'R_D}}\,,\hskip 1cm
  \vec \rho '={\vec r\over\sqrt{\lambda '\varepsilon_{A'}R_{A'}}}\,.
\end{equation}
The composition of the corresponding hyperbolic fractional Fourier transformations makes sense if $\vec\sigma '=\vec\sigma$, that is, if $\lambda '\varepsilon_D'R_D=\lambda \varepsilon_DR_D$, or
\begin{equation}
  {n'\over \varepsilon_D'}={n\over \varepsilon_D}\,,\end{equation}
or
\begin{equation}
  n'\frak{s'}{R_D-d\primespe\over d'\tanh\beta '}=n\frak{s}{R_D-d\over d\tanh\beta}\,.
\end{equation}
The spherical cap ${\cal A}'$ is the coherent image of ${\cal A}'$, if $\frak{s}\beta =\frak{s}'\beta '$, that is, if $\frak{s}'/\tanh\beta '=\frak{s}/\tanh\beta$, so that
\begin{equation}
  n'{R_D-d\primespe\over d'}=n{R_D-d\over d}\,,\label{eq158a}
\end{equation}
and, eventually, we obtain the conjugation formula
\begin{equation}
  {n'\over d\primespe}={n\over d}+{n'-n\over R_D}\,.\end{equation}

\subsubsection{Lateral magnification}

The beginning of Sect.\ \ref{sect515} remains valid and in particular Eq.\ (\ref{eq161}) also does.
 Since $\frak{s}\beta =\frak{s}'\beta '$, we have $\frak{s}\tanh\beta =\frak{s}'\tanh\beta '$, and Eq.\ (\ref{eq64}) is replaced with
  \begin{equation}
    {m_{\rm v}}^{\!2}={\lambda '\varepsilon_\basind{A'}R_{A'}\over\lambda \varepsilon_\basind{A}R_A}={\frak{s}'\over \frak{s}}\,{n\over n'}\,{d'\over d}{R_{A'}(R_A+d)\over R_A(R_{A'}+d')}\,{\tanh\beta '\over \tanh\beta}={nd\primespe\over n'd}{R_{A'}(R_A+d)\over R_A(R_{A'}+d\primespe)}\,.
  \end{equation}
  As in Sect.\ \ref{sect515}, we have
  \begin{equation}
    {R_AR_D\over (R_A+d)(R_D-d)}={R_DR_{A'}\over (R_D-d')(R_{A'}+d\primespe)}\,,\end{equation}
  so that, according to Eq.\ (\ref{eq158a})
  \begin{equation}
    {m_{\rm v}}^{\!2}={nd'\over n'd}\,{R_D-d\primespe\over R_D-d}= {n^2d\primespe^2\over n'^2d^2}\,.\label{eq198}\end{equation}
  Eventually, we have $-\beta d>0$ and $\beta 'd\primespe >0$, so that
    \begin{enumerate}
    \item[$\bullet$] if $\frak{s'}=\frak{s}$,  then $\beta '=\beta$ and  $d\primespe /d<0$.  Then $d$ and $d\primespe$ have opposite signs and the composition is given by Eq.\ (\ref{eq187}), that is, $\vec \rho '=-\vec \rho$, which means $m_{\rm v}<0$. 
    \item[$\bullet$]  if $\frak{s'}=-\frak{s}$,  then $\beta '=-\beta$ and $d\primespe /d>0$. Then $d$ and $d\primespe$ have the same sign and the composition is given  by Eq.\ (\ref{eq186}), so that $\vec\rho =\vec\rho '$, which means $m_{\rm v}>0$.
    \end{enumerate}
    We conclude that $m_{\rm v}$ has the sign of $d\primespe /d$, so that Eq.\ (\ref{eq198}) leads to
    \begin{equation}
   m_{\rm v}={nd\primespe\over n'd}\,.\end{equation}

\subsubsection{Conjugation of centers of curvature. Radius-magnification law}
  The results and proofs are exactly those of Sects.\ \ref{sect413} and \ref{sect414}.

\subsubsection{Field transfer from ${\cal A}$ to its image ${\cal A}'$}

We remark that both Eqs.\ (\ref{eq178}) and (\ref{eq179}) are still valid. As in Sect.\ \ref{sect518}, we conclude
\begin{equation}
  U_{A'}(\vec r')={1\over m_{\rm v}}\,U_A\left({\vec r'\over m_{\rm v}}\right)\,.\end{equation}

\subsection{Extension to imaging by a centered system}
The results of Sect.\ \ref{sect51} and \ref{sect52} are exactly those mentioned at the beginning of Sect.\ \ref{sect5}.  They can also be obtained by  composing two circular fractional-order Fourier transformations, when the geometrical configuration is appropriate \cite{PPF3,PPF1}.

If the refracting surface ${\cal D}$ is a plane, then $R_D$ is infinite, so that Eq.\ (\ref{eq140}) becomes
\begin{equation}
  {n'\over d\primespe}={n\over d}\,,\end{equation}
which is the conjugation formula for the refracting plane. Then $m_{\rm v}=1=m_{\rm c}$, regardless of the position of the object, so that and $m_{\rm r}=n'/n$.

The conjugation formula for the spherical cap is usually written in the form
\begin{equation}
  {n'\over d\primespe}={n\over d}+{n'\over f'}\,,\label{eq206}\end{equation}
where $f'=n'R_D/(n'-n)$ is the image focal-length of the refracting cap.

The previous findings also hold for a spherical mirror. Since a centered system is a sequence of refracting spherical caps or mirrors, it can be shown that those findings are valid for an objective lens (that is, a centered system with foci) \cite{PPF1}. If $H$ and $H'$ are the principal (or unit) points (on the axis) of the considered objective lens, the conjugation formula (\ref{eq206})  holds  with $d=\overline{HV_A}$, $d\primespe =\overline{H'V_{A'}}$, and $f'=\overline{H'F'}$, where $F'$ denotes the image focus. 
Apart from the conjugation formula, the previous results are valid for an afocal system \cite{PPF1}.

\section{Conclusion}\label{conc}

The use of hyperbolic fractional-order Fourier transformations completes the mathematical representation of diffraction by circular fractional Fourier transformations, in the framework of fractional Fourier optics.  A fractional-order Fourier transfor\-ma\-tion---circular or hyperbolic---can describe the field transfer from an emitter to a receiver, whatever the distance between them  and  their radii of curvature.

The compositions of two hyperbolic fractional transformations of the same kind lead to establish the properties of coherent imaging by a refracting spherical cap. They extend and complete  the  approach based on composing two circular transformations.  They also illustrate the method of the fractional Fourier transformation  in diffraction theory, according to which consistent results are obtained by only manipulating the orders of the transformations, without resorting to their explicit integral writings.

The previous findings can be applied to the theory of optical resonators, specifically for describing the field transfers between the resonator  mirrors \cite{PPF4,PPF5,PPF6}. Stable resonators correspond to field transfers represented by circular fractional transformations. 
Unstable resonators correspond to hyperbolic transformations and fall into two categories, depending on whether the  transformations involved are of the first or the second kind \cite{PPF5,PPF6}.

\appendix

\section{Proof of ${\cal H}_0={\cal I}$}\label{appenA}

We assume $\beta \ne 0$ and will prove that ${\cal H}_\beta$ tends to ${\cal I}$ when $\beta$ tends to zero.

\medskip
\noindent {\its i.} Let $\vec \rho =(x,y)$, $\vec \rho '=(x',y')$ and $\vec\eta =(u,v)$ 
  be three vectors belonging to ${\mathbb R}^2$.  
The scalar product of $\vec \rho$ and $\vec \rho \primespe$ is $\vec\rho \vec\cdot\vec \rho \primespe=x\,x'+y\,y'$ and similarly for the other vectors ($\vec\rho\vec\cdot\vec\eta= xu+yv$, etc.).

Let $S_\beta$ be defined on ${\mathbb R}^2\times {\mathbb R}^2$ by
\begin{equation}
  S_\beta(\vec \rho \primespe,\vec \eta )={\|\vec \rho \primespe\|^2-\|\vec \eta \|^2\over 2}\tanh\beta +{\vec\eta \vec\cdot\vec \rho \primespe\over \cosh\beta}\,,\label{eq184}
\end{equation}
and $\Phi_\beta$ by
\begin{equation}\Phi_\beta(\vec\rho ,\vec\rho\primespe)=-{(\|\vec\rho\|^2+\|\vec\rho \primespe\|^2) \cosh\beta-2\vec \rho \vec\cdot\vec\rho\primespe\over \sinh\beta}\,.
\end{equation}
(The function $S_\beta$ is known as the generating function \cite{Ego,Hor4,Nou} of the operator ${\cal U}_\beta$ that will be defined later on.)

\medskip
\noindent{\its ii.} Let $\Psi_\beta$ be defined by
\begin{equation}\Psi_\beta (\vec\rho, \vec \rho ',\vec \eta) ={\tanh\beta\over 2}\,\left\|\vec\eta  +{\vec\rho \cosh\beta -\vec \rho \primespe\over \sinh\beta}\right\|^2\,.
  \end{equation}
We have then
\begin{equation}
  S_\beta(\vec \rho \primespe,\vec \eta )-\vec\eta\vec\cdot\vec\rho=-\Psi_\beta(\vec\rho, \vec \rho \primespe,\vec \eta)-{1\over 2}\,\Phi_\beta (\vec\rho ,\vec\rho \primespe)\,.\label{eq203}
\end{equation}
A proof is as follows. We derive
\begin{eqnarray}
\Psi_\beta(\vec\rho ,\vec\rho \primespe,\vec\eta )+{1\over 2}\Phi_\beta (\vec \rho,\vec \rho \primespe)\rap &=&\rap {\tanh\beta\over 2}\left\|\vec\eta+{\vec\rho\cosh\beta -\vec\rho \primespe\over\sinh\beta}\right\|^2\! -{(\|\vec \rho \primespe\|^2+\|\vec \rho\|^2)\cosh\beta -2\vec \rho \vec\cdot  \vec \rho \primespe\over \sinh\beta}  \nonumber \\
&=&\rap
{\|\vec \eta\|^2\over 2}\tanh\beta +{\|\vec\rho\|^2\cosh\beta\over 2\sinh\beta}+{\|\vec\rho \primespe\|^2\over 2\cosh\beta\sinh\beta}-{\vec \rho\vec \cdot\vec \rho\primespe\over \sinh\beta}+\vec\eta\vec\cdot\vec \rho \nonumber\\
& &\hskip 1.5cm -{\vec\eta\vec\cdot\vec \rho\primespe\over \cosh\beta} -{\|\vec\rho\|^2\over 2}\,{\cosh\beta\over \sinh\beta}-{\|\vec\rho \primespe\|^2\over 2}\,{\cosh\beta\over \sinh\beta}+{\vec \rho \vec\cdot\vec \rho\primespe\over \sinh\beta} \nonumber\\
&=&\rap{\|\vec \eta\|^2\over 2}\tanh\beta+{\|\vec \rho \primespe\|^2\over 2\sinh\beta}\left({1\over \cosh\beta}-\cosh\beta \right)+\vec\eta\vec\cdot\vec \rho-{\vec\eta\vec\cdot\vec \rho \primespe\over\cosh\beta} \nonumber\\
&=&\rap {\|\vec\eta\|^2\over 2}\tanh\beta -{\|\vec \rho \primespe\|^2\over 2}\tanh\beta -{\vec\eta\vec\cdot\vec \rho '\over \cosh\beta}+\vec\eta\vec\cdot\vec \rho\nonumber \\
&=&\rap -{\|\vec\rho \primespe\|^2-\|\vec \eta\|^2\over 2}\tanh\beta  -{\vec\eta\vec\cdot\vec \rho '\over \cosh\beta}+\vec\eta\vec\cdot\vec \rho\nonumber \\
&=&  -S_\beta(\vec \rho \primespe,\vec \eta )+\vec\eta\vec\cdot\vec\rho\,,
\end{eqnarray}
and we obtain Eq.\ (\ref{eq203}).

\medskip
\noindent{\its iii.} If $\widehat f$ denotes the Fourier transform of $f$, we define an operator ${\cal U}_\beta$ on ${\cal S}({\mathbb R}^2)$ (the space of rapidly decreasing functions) by
\begin{equation}
  {\cal U}_\beta [f](\vec \rho \primespe)={1\over \cosh \beta}\int_{{\mathbb R}^2}\E^{-2\I\pi S_\beta (\vec\rho ', \vec \eta )}\,\widehat f(\vec \eta )\,\D\vec \eta\,.\label{eq188}\end{equation}
By the Fubini-Tonelli theorem we obtain
\begin{eqnarray}
  {\cal U}_\beta [f](\vec \rho \primespe)\rap &=&\rap {1\over \cosh \beta}\int_{{\mathbb R}^2}\E^{-2\I\pi S_\beta (\vec\rho ', \vec \eta )}
  \left(\int_{{\mathbb R}^2}\E^{2\I\pi \vec\rho \vec\cdot\vec\eta}\,f(\vec \rho)\,\D\vec \rho\right)\,\D\vec \eta\nonumber \\
  &=&\rap {1\over \cosh \beta}\int_{{\mathbb R}^2}f(\vec \rho)\,\left(\int_{{\mathbb R}^2} \E^{-2\I\pi [S_\beta (\vec\rho ', \vec \eta ) - \vec\rho \vec\cdot\vec\eta ]}\,\D\vec\eta\right)\,\D\vec \rho \nonumber \\
  &=&\rap  {1\over \cosh \beta}\int_{{\mathbb R}^2}\E^{\I\pi \Phi_\beta (\vec \rho ,\vec\rho ')} f(\vec \rho)\,I(\vec \rho ,\vec \rho \primespe )\,\D\vec \rho\,, 
\end{eqnarray}
where
\begin{equation}
  I(\vec \rho ,\vec \rho \primespe)=\int_{{\mathbb R}^2} \E^{2\I\pi \Psi_\beta(\vec\rho, \vec \rho ',\vec \eta)}\,\D\vec\eta
  =\int_{{\mathbb R}^2} \exp\left(\I\pi \tanh\beta \left\|\vec\eta  +{\vec\rho \cosh\beta -\vec \rho \primespe\over \sinh\beta} \right\|^2\right)\D\vec\eta\,.
\end{equation}
We derive
\begin{eqnarray}
I(\vec \rho,\vec \rho \primespe)\rap &=&\rap \exp\left({\I\pi \|\vec \rho\cosh\beta -\vec \rho \primespe\|^2\over \cosh\beta\sinh\beta }\right)\nonumber \\
& &\hskip 1cm \times \;\;\int_{\mathbb R}\!\!\exp(\I\pi \|\vec\eta\|^2\tanh\beta)\exp\left[2\I\pi \vec\eta \,\vec\cdot\left(\vec \rho -{\vec \rho \primespe\over\cosh\beta}\right)\right]\D\vec \eta\,.\nonumber \\
&=&\rap \exp\left(\I\pi \left\|\vec \rho  -{\vec \rho \primespe\over\cosh\beta}\right\|^2\coth\beta\right)\nonumber \\
& &\hskip 1cm \times \;\;\int_{\mathbb R}\!\!\exp(\I\pi  \|\vec\eta\|^2\tanh\beta)\exp\left[2\I\pi \vec\eta\,\vec\cdot \left(\vec \rho -{\vec \rho \primespe\over\cosh\beta}\right)\right]\,\D\vec \eta\,.\label{eq191}
\end{eqnarray}
We use
\begin{equation}
{1\over \I A}\exp \left({\I\pi\over A} \|\vec\eta\|^2\right)\;\rightleftharpoons\; \exp (-\I\pi A  \|\vec\nu\|^2)\,,
\end{equation}
where $\vec \nu$ denotes the conjugated  variable of $\vec \eta$. Here, with $A=1/\tanh\beta$ ($\tanh\beta \ne 0$), we obtain
\begin{equation}
\exp (\I\pi  \|\vec\eta\|^2\tanh\beta )\;\rightleftharpoons \; {\I\over \tanh\beta }\exp \left(-{\I\pi  \|\vec\nu\|^2\over \tanh\beta}\right)\,,\end{equation}
then
\begin{eqnarray}
& &\int_{\mathbb R}\!\!\exp(\I\pi \|\vec\eta\|^2\tanh\beta)\exp\left[2\I\pi \vec\eta\vec\cdot \left(\vec \rho -{\vec \rho\primespe\over\cosh\beta}\right)\right]\,\D\vec\eta \nonumber \\
& & \hskip 3cm ={\I \over \tanh\beta }\,\exp \left(-{\I\pi \over \tanh\beta} \left\|\vec \rho -{\vec \rho\primespe\over\cosh\beta}\right\|^2\right)\nonumber \\
& & \hskip 3cm ={\I\over  \tanh\beta}\, \exp \left(-{\I\pi} \left\|\vec \rho -{\vec \rho\primespe\over\cosh\beta}\right\|^2\coth\beta\right)\,,
\end{eqnarray}
and eventually
\begin{equation}
I(\vec \rho,\vec \rho \primespe)={\I\over \tanh\beta }\,.\end{equation} 
We conclude
\begin{eqnarray}
{\cal U}_\beta [f](\vec\rho \primespe)\rap &=&\rap {\I\over \sinh\beta}\int_{{\mathbb R}^2}\E^{\I\pi \phi_\beta (\vec \rho,\vec \rho ')}\,f(\vec \rho)\,\D \vec \rho\,,\nonumber \\
&=&\rap 
{\I\over \sinh\beta }\,\E^{-\I\pi \|\vec\rho '\|^2\coth\beta}
\int_{{\mathbb R}^2}\E^{-\I\pi \|\vec\rho\|^2\coth\beta}\exp\left({2\I\pi \vec\rho '\vec\cdot\vec\rho\over\sinh\beta}\right)\,f(\vec\rho)\,\D \vec\rho \,,\end{eqnarray}
so that
\begin{equation}
  {\cal H}_\beta =\E^{\beta }\,{\cal U}_\beta \,.\label{eq197}\end{equation}

\medskip
\noindent{\its iv.}
According to Eq.\ (\ref{eq184}), when  $\beta$ tends to $0$, we obtain
\begin{equation}
  S_0(\vec\rho \primespe,\vec\eta )=\vec\eta\vec\cdot\vec\rho \primespe\,,\end{equation}
and Eq.\ (\ref{eq188}) becomes
\begin{equation}
   {\cal U}_0 [f](\vec \rho \primespe)=\int_{{\mathbb R}^2}\E^{-2\I\pi \vec\eta\vec\cdot\vec\rho '}\,\widehat f(\vec \eta )\,\D\vec \eta =f(\vec\rho \primespe)\,,\end{equation}
which holds for every function in ${\cal S}({\mathbb R}^2)$ and means that ${\cal U}_0={\cal I}$. From Eq.\ (\ref{eq197}), we deduce ${\cal H}_0={\cal I}$.
(The result is extended to tempered distributions by using $\langle {\cal U}_0[T],f\rangle = \langle T,{\cal U}_0[f]\rangle = \langle T,f\rangle $.) \qed


\section*{Acknowledgements} The author thanks Prof.\ Jean-Fran\c cois Marini for his encouragement during the preparation of this article.



\begin{thebibliography}{references}
\bibitem{PPF7} P. Pellat-Finet, ``Fractional-order Fourier transformations as pseudo-differential operators. Applications to diffraction and imaging by optical systems.'' To be published.
\end{thebibliography}
\end{document}